\documentclass[12pts, reqno]{article}
\usepackage{amsmath,amsfonts,amssymb,amscd,amsbsy, color}
\usepackage[dvips]{graphicx}
\usepackage{comment}
\usepackage{courier}
\usepackage{fancybox}
\usepackage{multimedia}
\usepackage{graphicx,epstopdf}
\usepackage{appendix}
\usepackage{wasysym}
\usepackage{url}
\usepackage{natbib}
\usepackage[maxfloats=100]{morefloats}

\newcommand\beq[1]{ \begin{equation}\label{#1} }
\newcommand{\eeq}{ \end{equation} }
\newcommand{\beqno}{ \[ }
\newcommand{\eeqno}{ \] }
\newcommand\beqa[1]{ \begin{eqnarray} \label{#1}}

\newcommand{\eeqa}{ \end{eqnarray} }
\newcommand{\beqano}{ \begin{eqnarray*} }
\newcommand{\eeqano}{ \end{eqnarray*} }
\newcommand\equ[1]{{\rm (\ref{#1})}}

\def\G{{\mathcal G}}

\begin{document}

\title{Poynting-Robertson drag and solar wind in the space debris problem}

\author{
Christoph Lhotka\\
{\small Space Research Institute, Austrian Academy of Sciences,}\\
{\small Schmiedlstrasse 6, 8042 Graz (Austria)}\\
{\small \rm christoph.lhotka@oeaw.ac.at}
\and
Alessandra Celletti\\
{\small Department of Mathematics, University of Roma Tor Vergata,}\\
{\small Via della Ricerca Scientifica 1, 00133 Roma (Italy)}\\
{\small \rm celletti@mat.uniroma2.it}
\and
C\u at\u alin Gale\c s\\
{\small Department of Mathematics, Al. I. Cuza University,}\\
{\small Bd. Carol I 11, 700506 Iasi (Romania)}\\
{\small \rm cgales@uaic.ro}
}

\date{Received: date / Accepted: date}
\maketitle

\label{firstpage}

\begin{abstract}
We analyze the combined effect of Poynting-Robertson and solar wind drag on space debris.
We derive a model within Cartesian, Gaussian and Hamiltonian frameworks.
We focus on the geosynchronous resonance, although the results can be easily generalized to any resonance.
By numerical and analytical techniques, we compute the drift in semi-major axis due to Poynting-Robertson
and solar wind drag. After a linear stability analysis of the equilibria,
we combine a careful investigation of the regular, resonant,
chaotic behavior of the phase space with a long-term propagation of a sample of initial
conditions. The results strongly depend on the value of the area-to-mass ratio of the debris,
which might show different dynamical behaviors: temporary capture or escape from the geosynchronous resonance, as well as
temporary capture or escape from secondary resonances involving the rate
of variation of the longitude of the Sun. Such analysis shows that Poynting-Robertson
and solar wind drag must be taken into account, when looking at the long-term behavior
of space debris. Trapping or escape from the resonance can be used to place the debris
in convenient regions of the phase space.
\end{abstract}

{\bf keywords}
Poynting-Robertson effect, Solar wind, Geostationary orbit, Space debris

\section{Introduction}
The population of space debris shows a wide variety of different
case studies: micro-metric particles to meter-size debris, small
to large area-to-mass ratios, circular to highly inclined orbits,
etc. These characteristics of the space debris, together with
their actual location in LEO, MEO, GEO\footnote{LEO stands for \sl
Low Earth Orbit \rm ranging from 90 to 2\,000 km of altitude, MEO
stands for Medium Earth Orbit running between 2\,000 and 30\,000
km of altitude, GEO stands for Geostationary Earth Orbit at about
35\,786 km of altitude.}, lead to consider several models as well
as a different hierarchy of the forces which contribute to shape
the dynamics. For example, it was widely shown (see, e.g.,
\cite{CG2014b}, \cite{Kuz2011}, \cite{Valketal2009} and references
therein) that the effect of solar radiation pressure on GEO and
MEO objects is more relevant for larger area-to-mass ratios. The
dissipative contribution due to Poynting-Robertson and solar wind
is definitely considered much less important. However, such
dissipative effects might become relevant on micro-meter size
particles as well on large area-to-mass ratio space debris
(various objects with high area-to-mass ratios are described,
e.g., in \citet{FruSch2012}).

The aim of the current study is to settle the question of the role of
Poynting-Robertson and solar wind (hereafter PR/SW) drag on space debris dynamics.
The relevance of this question stems from the consideration that the drag might provoke
a drift of the debris towards space regions where operating satellites are placed.
To avoid collisions with functional satellites and a consequent possible generation
of further debris, it is crucial to have a full control of the dynamics of space debris,
including the prediction over long time scales of minor, but still relevant, effects
like PR/SW.
Our study shows that not only Poynting-Robertson drag, but also
solar wind drag, are prominent forces that need to be included in the model to get an
accurate estimate of the drift of space debris in the near-Earth environment.

The first studies on the Poynting-Robertson effect in the artificial satellite
problem date back to \citet{Sla1980, Sla1983}, where the author states that the
semi-major axis of a near synchronous satellite with small area-to-mass ratio
can decrease at a rate of about 1 $mt/yr$.
The secular evolution of geostationary objects
caused by light pressure alone has been treated in
\citet{SmiMik1993,SmiMik1995}. Numerical simulations in the neighborhood of
the 1:1, 1:2, and 1:3 resonances can be found in \citet{KuzEtAl2012,
KuzEtAl2013}. The authors estimate the secular effect caused by the
Poynting-Robertson drag for various area-to-mass ratios being of the order of
hundreds of meters per year, approximately. In \citet{KuzEtAl2014} the authors
provide a series of numerically obtained estimates of drift rates for various
high order resonances close to the geosynchronous orbit. They find drift rates
(in absolute magnitudes) ranging from about 29 $mt/yr$ (9:11 resonance) to about
142 $mt/yr$ (5:4 resonance) with a variation of 33 $mt/yr$ to 75 $mt/yr$ close to the
geosynchronous orbit (see Table 2 in \citet{KuzEtAl2014}). Secular rates of drift in semi-major axis
of about 500 $mt/yr$ have also numerically been estimated in \citet{KuzZak2015}
for high area-to-mass ratio objects in highly elliptical orbits, the so-called Molniya orbits,
close to the 22:45 resonance.

As it has been recognized in \citet{Kuz2011}, drift rates in the vicinity of the
geosynchronous orbit may differ by orders of magnitude. Typical estimates for
standard area-to-mass ratios range from -23 $km/yr$ \citep[][]{SmiMik1993}, -59 $mt/yr$
\citep[][]{Sla1980}, -51 $mt/yr$ \citep[][]{TueAvd2006}, and -80 $mt/yr$
\citep[all values taken from][]{Kuz2011}.  The divergence is
clearly due to resonance effects, so that the drift in semi-major axis strongly
depends on the initial condition, i.e.  the distance from the exact resonance, that
itself becomes shifted for large area-to-mass ratios.

The aim of our study is to provide a detailed investigation of the drift in
semi-major axis of space debris of high area-to-mass ratios, subject to
the lower degree gravitational field of the Earth, the gravitational attraction
due to the Moon and the Sun, direct solar radiation pressure, the
Poynting-Robertson drag and the solar wind drag forces. First, we provide the model
in different frameworks: Cartesian coordinates, Gaussian equations of motion and using
a Hamiltonian approach.
We investigate the dynamics
by means of secular perturbation theory, and isolate the effects due to
PR/SW-drag from additional perturbations of Kepler's orbit. We provide
realistic estimates of the drift by means of simple formulae and check our
results by means of a numerical integration of the full problem.

On the basis of a newly developed, fully non-resonant secular theory, our estimate of the drift rate
outside the geosynchronous resonant regime of motion turns out to be of the order of 40 $mt/yr$,
a value which is confirmed by numerical experiments.
Averaging over the fast variables, one obtains a simple system of equations which allows us to
compute the location of the equilibria for the geostationary orbit, within different models, possibly including
Poynting-Robertson effect and solar wind.
Under the influence of dissipative forces, we find different behaviors in the neighborhood of the
geosynchronous resonance: we find temporary capture, with a transient time varying according to
the orbital properties and to the value of the area-to-mass ratio; we observe escape orbits or
rather transitions between the main resonance and other resonances involving the rate of variation
of the longitude of the Sun.
Our investigation includes a detailed study of possible chaotic motions at the border
of the resonance; such regions might be used to transfer the debris without much effort,
but rather relying on the dynamical properties of the chaotic regions.

We mention that to get a comprehensive description of the dynamics, other effects should have been considered.
Among the others, we mention the long-term
periodic evolution of space debris trajectories caused by successive Earth's shadow
crossings, which has been investigated, e.g., in \citet{Hub2013}; a detailed model of
the so-called BYORP effect, which has been derived in \citet{MahSch2010}.

This paper is organized as follows. In Section~\ref{model}
we present the dynamical model we are going to use, and derive the geocentric
formulation of PR/SW-drag. A qualitative discussion
of the PR/SW-drag on artificial satellite motion can be found in Section~\ref{qua}.
The detailed investigation of the dynamics close to the resonant regimes of motion
is provided in Section~\ref{num}. A discussion of the
results can be found in Section~\ref{sec:conclusions}.

\section{The dynamical model}
\label{model}

In this Section we present a model describing the motion of a spherical space debris object (hereafter SDO)
subject to the gravitational influence of the Earth, the gravitational
attraction of the Moon, and the Sun, solar radiation pressure, solar wind,  and
Poynting-Robertson drag force. The Cartesian formulation is given in Section~\ref{subsec:cartesian},
the Gaussian equations of motion are presented in Section~\ref{subsec:Gauss_form},
while a Hamiltonian description is given in Section~\ref{subsec:near_Hamiltonian}.

\subsection{Cartesian framework} \label{subsec:cartesian}

Let $\vec r$ be the position of the SDO of mass $m$ in a quasi-inertial, geocentric
reference frame, $\mathbf e=(\vec{e}_1,\vec{e}_2,\vec{e}_3)$, and denote by
$\vec r_M$, $\vec r_S$ the position vectors of the Moon, and the Sun,
respectively. Then, the equation of motion of the SDO is given by:
\beq{eq1}
\frac{d^2\vec{r}}{dt^2}=
-\frac{d}{d\vec{r}}\bigg[
V_E\left(\vec{r}\right)
+ V_M\left(\vec{r}, \vec{r}_M\right)
+ V_S\left(\vec{r}, \vec{r}_S\right)
+ V_{SRP}\left(\vec{r}, \vec{r}_S\right)\bigg]
+ \vec{F}_{PR/SW}\left(\vec{r},\vec{r}_S\right) \ ,
\eeq
where $V_E$, $V_M$, $V_S$ are the gravitational potentials of the Earth, the
Moon, and the Sun, $V_{SRP}$ is the potential of solar radiation pressure, and
$\vec{F}_{PR/SW}$ labels the combined Poynting-Robertson and solar wind drag terms,
respectively.  Let $\mu=\G m_E$ be the geocentric gravitational constant with
constant of gravity $\G$, and mass of the Earth $m_E$. The geopotential $V_E$ is
given in a synodic reference frame, with unit vectors $\mathbf f=(\vec{f}_1,
\vec{f}_2, \vec{f}_3)$, and rotating with the same angular velocity of the
Earth, as (\cite{MonGil2000}):
$$
V_E(r,\phi,\lambda)={{\G m_E}\over r}\
\sum_{n=0}^\infty \Big({R_E\over r}\Big)^n\ \sum_{m=0}^n P_{nm}(\sin\phi)\
(C_{nm}\cos m\lambda+ S_{nm}\sin m\lambda)\ .
$$

Here $r$, $\phi, \lambda$ are Earth-fixed spherical coordinates (radius, co-latitude,
and longitude with $\lambda=0$ corresponding to the Greenwich mean meridian), and $R_E$,
$C_{nm}$, $S_{nm}$ are the mean equatorial radius of the Earth and the (not normalized) Stokes coefficients that
enter the spherical expansion of the Earth's gravitational field up to degree
$n$ and order $m$. The quantities $P_{nm}$ are the associated Legendre polynomials
(in the geophysical sense, see Appendix A).  We denote by $m_M$, $m_S$ the masses
of the Moon and the Sun, respectively. The gravitational potentials due to the
third-body, point-mass like interactions take the functional form (\cite{Mur1999}):
$$
V_{k}\left(\vec r,\vec r_k\right) = \G m_k\left(\frac{1}{\|\vec r-\vec r_k\|} -
\frac{\vec r \cdot \vec r_k}{r_k^3}\right) \ ,
$$
with index $k=M$ for the Moon and $k=S$ for the Sun. We are left to provide $V_{SRP}$ and $F_{PR/SW}$. Let $\vec X$,
$\vec V$, $R$ denote the position, velocity, and heliocentric distance of
the SDO in the heliocentric frame of reference $\mathbf g=({\vec g}_1, {\vec
g}_2, {\vec g}_3)$, respectively. Furthermore, we denote by
${\hat g}_{R}=\vec{X}/{R}$ the radial unit
vector along the line connecting Sun - SDO. The standard definition of the
combined acceleration due to solar radiation pressure, Poynting-Robertson,
and solar wind drag force, say ${\vec F'}$, is given by (\cite{BLS1979,
Kla2014, LhoCel2015}):
\beq{RF}
\vec{F'}=\frac{\beta \G m_S}{{R}^2}\left[{\hat g}_{R}-\left(1+\frac{\eta}{Q}\right)
\left(
\frac{\vec V\cdot\hat{g}_{R}}{c} \ \hat{g}_{R}+\frac{\vec V}{c}
\right)\right] \ .
\eeq
Here, $\beta$ is the ratio of the magnitude of radiation force over solar gravitational
attraction (\cite{Koc2006}):
\beqno
\beta = \frac{SQA}{c}/\frac{\G mm_S}{{R}^2} \simeq 7.6\times10^{-4}
\ Q \ \frac{A \ [mt^2]}{m \ [kg]} \ ,
\eeqno
with energy flux constant $S$, spectrally averaged dimensionless efficiency
factor for radiation pressure $Q$, speed of light $c$, and the area cross
section $A$ and mass $m$ of the SDO, respectively. In addition,  $\eta$ in \equ{RF}
is the dimensionless solar wind drag efficiency factor, i.e. the ratio of solar
wind over Poynting-Robertson drag force (approximately about equal to $1/3$).

As a first step, we express \equ{RF} in the geocentric frame $\mathbf{e}$ by
making use of the relation $\vec{X}=\vec{r}-\vec{r}_S$:
\beq{RFgeo}
\vec{F'}=\beta \G m_S\ {{\vec{r}-\vec{r}_S}\over {|\vec{r}-\vec{r}_S|^3}}
-\frac{\beta \G m_S}{c}\left(1+\frac{\eta}{Q}\right)
\left\{ {{\vec{r}-\vec{r}_S}\over {|\vec{r}-\vec{r}_S|^3}}
(\vec{\dot r}-\vec{\dot r}_S)
+ {{\vec{\dot r}-\vec{\dot r}_S}\over {|\vec{r}-\vec{r}_S|^2}}\right\}\ .
\eeq
We notice that the first term of \equ{RFgeo} can be derived from the potential
$$
V_{SRP}\left(\vec r,\vec r_S\right)=\beta \G m_S\left(\frac{1}{|\vec r - \vec r_S |}\right) \ .
$$
For clarity of exposition we split the conservative contribution from
\equ{RFgeo}, and define the remaining force component ${\vec F}_{PR/SW}$ in
\equ{eq1} as:
\beq{SWPR}
{\vec F}_{PR/SW} = {\vec F'} - \frac{dV_{SRP}\left(\vec r, \vec r_S\right)}{d\vec r} \ .
\eeq

Denoting by $(x,y,z)$, $(x_M, y_M, z_M)$, $(x_S, y_S, z_S)$ the coordinates in
the quasi--inertial frame $\mathbf e$ of the SDO, Moon and Sun, respectively, the components of the equations of
motion are then simply given by:
\beqa{eq3}
\ddot x&=&V_x(x,y,z, \theta)-\G m_S\Bigl({{x-x_S}\over {|\vec{r}-\vec{r}_S|^3}}+{x_S\over r_S^3}\Bigr)-\G m_M \Bigl({{x-x_M}\over {|\vec{r}-\vec{r}_M|^3}}+{x_M\over r_M^3}\Bigr)
+\beta \G m_S\ {{x-x_S}\over {|\vec{r}-\vec{r}_S|^3}} \ - \nonumber \\
&&\frac{\beta \G m_S}{c}\left(1+\frac{\eta}{Q}\right)
\left\{ {{x-x_S}\over {|\vec{r}-\vec{r}_S|^3}}
(\dot x-\dot x_S)
+ {{\dot x-\dot x_S}\over {|\vec{r}-\vec{r}_S|^2}}\right\} \nonumber\\
\ddot y&=&V_y(x,y,z, \theta)-\G m_S\Bigl({{y-y_S}\over {|\vec{r}-\vec{r}_S|^3}}+{y_S\over r_S^3}\Bigr)-\G m_M \Bigl({{y-y_M}\over {|\vec{r}-\vec{r}_M|^3}}+{y_M\over r_M^3}\Bigr)
+\beta \G m_S\ {{y-y_S}\over {|\vec{r}-\vec{r}_S|^3}} \ - \nonumber \\
&&\frac{\beta \G m_S}{c}\left(1+\frac{\eta}{Q}\right)
\left\{ {{y-y_S}\over {|\vec{r}-\vec{r}_S|^3}}
(\dot y-\dot y_S)
+ {{\dot y-\dot y_S}\over {|\vec{r}-\vec{r}_S|^2}}\right\} \nonumber\\
\ddot z&=&V_z(x,y,z, \theta)-\G m_S\Bigl({{z-z_S}\over {|\vec{r}-\vec{r}_S|^3}}+{z_S\over r_S^3}\Bigr)-\G m_M \Bigl({{z-z_M}\over {|\vec{r}-\vec{r}_M|^3}}+{z_M\over r_M^3}\Bigr)
+\beta \G m_S\ {{z-z_S}\over {|\vec{r}-\vec{r}_S|^3}} \ - \nonumber \\
&&\frac{\beta \G m_S}{c}\left(1+\frac{\eta}{Q}\right)
\left\{ {{z-z_S}\over {|\vec{r}-\vec{r}_S|^3}}
(\dot z-\dot z_S)
+ {{\dot z-\dot z_S}\over {|\vec{r}-\vec{r}_S|^2}}\right\} \ ,
\eeqa
where $(V_x,V_y,V_z)$ represent the three components of the derivatives of the
geopotential in the sidereal reference frame $\mathbf e$ that depend,
additionally, on the sidereal time $t$ through the Greenwich Meridian angle $\theta$.
The system of equations \equ{eq3} serves as the
basis for our numerical study\footnote{We notice that the solar radiation pressure
terms in \equ{eq3} are equivalent to the usual definition of solar radiation pressure in the
artificial satellite problem given by (\cite{MonGil2000}):
\beqno
C_rP_ra_S^2\bigg(\frac{A}{m}\bigg)
{{\vec{r}-\vec{r}_S}\over {|\vec{r}-\vec{r}_S|^3}} \ ,
\eeqno
where $C_r$, $P_r$, $a_S$ are the reflectivity coefficient $C_r$, and the radiation pressure $P_r$
located at $a_S=1\ AU$, respectively.}.

\subsection{Gauss' form of the equations of motions}
\label{subsec:Gauss_form}

For the qualitative description of the secular dynamics of the SDO we will also
work in Kepler elements and Delaunay variables. Let $a$ be the semi-major axis,
$e$ the eccentricity, $i$ the inclination, $\omega$ be the argument of perihelion,
$\Omega$ the longitude of the ascending node, and $M$ the mean anomaly defined in the
quasi-inertial reference frame $\mathbf e$. In this setting, the norm of the orbital angular
momentum is given by $h=\sqrt{1-e^2}\sqrt{\mu a}$, Kepler's 3rd law is $\mu=n^2a^3$,
and $n$ is the mean motion of the SDO. We start with Gauss' form of the perturbed Kepler equations
of motion (\cite{Fitzpatrick}):
\beqa{gauss}
\frac{da}{dt} &=&
\frac{2ah}{\mu\left(1-e^2\right)}
\left[e\sin f {\mathfrak F}_R + \left(1 + e\cos f\right) {\mathfrak F}_T\right] \ , \\
\frac{de}{dt} &=&
\frac{h}{\mu}
\left[\sin f {\mathfrak F}_R + \left(\cos f + \cos E\right) {\mathfrak F}_T\right] \ ,
\nonumber  \\
\frac{di}{dt} &=&
\frac{\cos\left(\omega+f\right)r}{h} {\mathfrak F}_N , \ \nonumber \\
\frac{d\omega}{dt} &=&
-\frac{h}{\mu}\frac{1}{e}
\left[\cos f {\mathfrak F}_R-\left(\frac{2+e\cos f}{1+e cos f}\right)\sin f {\mathfrak F}_T\right]
-\frac{\cos i\sin\left(\omega+f\right)r {\mathfrak F}_N}{h \sin i} \ , \nonumber \\
\frac{d\Omega}{dt} &=&
\frac{\sin\left(\omega+f\right)r}{h \sin i} {\mathfrak F}_N \ , \nonumber \\
\frac{dM}{dt} &=& n +
\frac{h}{\mu}\frac{\sqrt{1-e^2}}{e}
\left[\left(\cos f-\frac{2e}{1-e^2}\frac{r}{a}\right) {\mathfrak F}_R
-\left(1+\frac{1}{1-e^2}\frac{r}{a}\right)\sin f {\mathfrak F}_T\right] \nonumber \ .
\eeqa
Here, $\mathfrak F_R$, $\mathfrak F_T$, $\mathfrak F_N$ are the radial, tangential, and
normal components of a generic perturbing force, decomposed in the form:
\beq{gaussbis}
{\mathfrak F}=\mathfrak F_R {\mathbf o}_R + \mathfrak F_T {\mathbf o}_T + \mathfrak F_N
{\mathbf o}_N  \ ,
\eeq
where ${\mathbf o}_R=(\cos f,\sin f,0)$, ${\mathbf o}_T=(-\sin f, \cos f, 0)$,
and ${\mathbf e}_N={\mathbf o}_R\times{\mathbf o}_T$ are the radial,
tangential, and normal unit vectors defined in the orbital reference frame
$\mathbf o$ centered at $\vec r$ with true anomaly $f$.  The transformation
between $\mathbf o$ and $\mathbf e$ is given in terms of the rotation
matrix\footnote{See Appendix A for the definition of the rotation matrices.}
\beqno
{\mathfrak
R}={\mathfrak R}_3(\Omega)\cdot{\mathfrak R}_1(i)\cdot{\mathfrak R}_3(\omega) \ .
\eeqno
In this setting, the force function $\mathfrak F$ in \equ{gaussbis} is related to
${\vec F}_{PR/SW}=F_x {\vec e}_1 + F_y {\vec e}_2 + F_z {\vec e}_3$ in
\equ{SWPR}, by the expressions (\cite{Mou1914}):
\beqano
\mathfrak F_R = \left(F_x,F_y,F_z\right) \cdot {\mathfrak R} \cdot {\mathbf o}_R \ , \
\mathfrak F_T = \left(F_x,F_y,F_z\right) \cdot {\mathfrak R} \cdot {\mathbf o}_T \ , \
\mathfrak F_N = \left(F_x,F_y,F_z\right) \cdot {\mathfrak R} \cdot {\mathbf o}_N \ .
\eeqano

We remark, that using well known formulae for Taylor series expansions in the
two-body problem (see, e.g., \cite{mybook}), the right hand sides of \equ{gauss} can
be written in terms of the orbital elements of the SDO and the Sun, only\footnote{In addition,
all other perturbations in \equ{eq1} can be identified with $\mathfrak F$ in \equ{gauss} in
a straightforward way.}.

\subsection{Near Hamiltonian form}
\label{subsec:near_Hamiltonian}

Let $L=\sqrt{\mu a}$, $G=L\sqrt{1-e^2}$, $H=G\cos i$, $l=M$, $g=\omega$, $h=\Omega$
denote the Delaunay variables of the SDO. The evolution in time of these action-angle
like variables can be computed from \equ{gauss} and take the following expression:
\beqa{HAM}
\frac{dL}{dt}&=&\frac{\mu}{2L}\frac{da}{dt}
= -\frac{d\mathfrak H}{dl} + {\mathfrak f}_L
 \ , \nonumber \\
\frac{dG}{dt}&=&\frac{\mu G}{2L^2}\frac{da}{dt} - \frac{L\sqrt{L^2-G^2}}{G}\frac{de}{dt}
= -\frac{d\mathfrak H}{dg} + {\mathfrak f}_G
 \ , \nonumber \\
\frac{dH}{dt}&=&-\frac{\mu H}{2L^2}\frac{da}{dt} + \frac{HL\sqrt{L^2-G^2}}{G}\frac{de}{dt}
-\sqrt{G^2-H^2}\frac{di}{dt}
=-\frac{d\mathfrak H}{dh} + {\mathfrak f}_H
 \ ,  \nonumber \\
\frac{dl}{dt}&=&\frac{dM}{dt}
= \frac{d\mathfrak H}{dL} + {\mathfrak f}_l
 \ , \nonumber \\
\frac{dg}{dt}&=&\frac{d\omega}{dt}
= \frac{d\mathfrak H}{dG} + {\mathfrak f}_g
 \ , \nonumber  \\
\frac{dh}{dt}&=&\frac{d\Omega}{dt}
= \frac{d\mathfrak H}{dH} + {\mathfrak f}_h
 \ .
\eeqa
Here, ${\mathfrak H}$ denotes the Hamiltonian part of \equ{eq1} (that can be derived from
the potential terms):
$$
{\mathfrak H}=-{\mu_E^2\over {2L^2}}+V_E+V_M+V_S+V_{SRP}\ ;
$$
moreover, we have introduced the functions ${\mathfrak f}_L$, ${\mathfrak f}_G$,
${\mathfrak f}_H$, ${\mathfrak f}_l$, ${\mathfrak f}_g$, ${\mathfrak f}_h$, which stem from
the non-conservative contributions ${\vec F}_{PR/SW}$ in \equ{eq1}.

\section{Poynting-Robertson and solar wind drag}
\label{qua}

Based on the Gaussian equations given in \equ{gauss}, we proceed to investigate the effect of the
drag by averaging the equations of motion. A careful analysis provides a simple formula for the
drift rate of the semimajor axis, as well as the location and stability of the equilibrium
positions.

\subsection{Drift rate of the semimajor axis}
Let $a_S$, $e_S$, $i_S$, $\omega_S$, $\Omega_S$, $M_S$ be the orbital elements of the Sun
given in the reference frame $\mathbf e$. In this setting, the system \equ{gauss},
with $\mathfrak F$ derived on the basis of ${\vec F}_{PR/SW}$ alone, becomes:

\beqa{PRgauss}
\frac{da}{dt} &=&
-\frac{\G m_S}{a_S}\frac{\beta}{c}\left(1 + \frac{\eta}{Q}\right)
\left\{
 c_{\mathbf 0}^{(1)} + [1] + \frac{n_S}{n}\left(c_{\mathbf 0}^{(2)} + [2]\right)
\right\} \ ,
\nonumber \\
\frac{de}{dt} &=&
-\frac{\G m_S}{a_S^2}\frac{\beta}{c}\left(1 + \frac{\eta}{Q}\right)
\left\{
[3] + \frac{n_S}{n}\left(c_{\mathbf 0}^{(4)} + [4]\right)
\right\} \ ,
\nonumber \\
\frac{di}{dt} &=&
-\frac{\G m_S}{a_S^2}\frac{\beta}{c}\left(1 + \frac{\eta}{Q}\right)
\left\{
[5] + \frac{n_S}{n}\left(c_{\mathbf 0}^{(6)} + [6]\right)
\right\} \ ,
\eeqa
and
\beqa{PRgauss2}
\frac{d\omega}{dt} &=&
\frac{\G m_S}{a_S^2e\sin i}\frac{\beta}{c}\left(1+\frac{\eta}{Q}\right)
\left\{[7]+\frac{n_S}{n}[8]\right\} \ ,
\nonumber \\
\frac{d\Omega}{dt} &=&
\frac{\G m_S}{a_S^2\sin i}\frac{\beta}{c}\left(1+\frac{\eta}{Q}\right)
\left\{[9]+\frac{n_S}{n}[10]\right\} \ ,
\nonumber \\
\frac{dM}{dt} &=& n +
\frac{\G m_S}{a_S^2e}\frac{\beta}{c}\left(1+\frac{\eta}{Q}\right)
\left\{[11]+\frac{n_S}{n}[12]\right\} \ .
\eeqa
Here, the terms $[\#]$ in \equ{PRgauss}, \equ{PRgauss2} are the superposition of
periodic functions with wave number $\mathbf k$, and amplitude equal to
$c_{{\mathbf k},j}^{(\#)}$, $s_{{\mathbf k},j}^{(\#)}$, being themselves
polynomials in $e$, $e_S$, $\cos i$, $\cos i_S$, $\sin i$, $\sin i_S$,
respectively\footnote{Second order expansions in the small parameters $e$ and $e_S$ can be
obtained from the authors as Mathematica notebooks.}:
\beqano
\left[\#\right]=\sum_{j}
\left(\frac{a}{a_S}\right)^j
\bigg\{
\sum_{{\mathbf k}\in{\mathbb Z}^6} \hskip -.1in
&c_{{\mathbf k},j}^{(\#)}\big(e,e_S,i,i_S\big)
\cos\left(k_1 M+k_2 \omega+k_3\Omega+k_4 M_S + k_5 \omega_S + k_6\Omega_S \right)&  \ +
\nonumber \\
&s_{{\mathbf k},j}^{(\#)}\big(e,e_S,i,i_S\big)
\sin\left(k_1 M+k_2 \omega+k_3\Omega+k_4 M_S + k_5 \omega_S + k_6\Omega_S \right)& \hskip -.1in
\bigg\} \ .
\eeqano
On long time scales these terms average out with respect to the constant terms
$c_{{\mathbf 0},j}^{(\#)}$. Therefore, by setting $[\#]=0$ in \equ{PRgauss} we are left with
the following secular system:
\beqa{PRgaussSEC}
\frac{da}{dt} &=& -\frac{a}{a_S}\frac{2\G m_S}{a_S}\frac{\beta}{c}\left(1+\frac{\eta}{Q}\right)
\left[1+\frac{e_S^2}{2}-\cos i \cos i_S\left(1-\frac{e^2}{2}+\frac{5e_S^2}{2}\right)
\frac{n_S}{n}\right] \ , \nonumber\\
\frac{de}{dt} &=&
-\frac{n_S}{n}\frac{5\G m_S}{2a_S^2}\frac{\beta}{c}\left(1+\frac{\eta}{Q}\right)
e \cos i \cos i_S \ , \nonumber\\
\frac{di}{dt} &=&
-\frac{n_S}{n}\frac{\G m_S}{2a_S^2}\frac{\beta}{c}\left(1+\frac{\eta}{Q}\right)
\sin i\cos i_S\left(1+2e^2+\frac{5e_S^2}{2}\right) \ ,
\eeqa
while the time derivatives of the angles reduce to $dM/dt=n$, and $d\omega/d
t=d\Omega/d t=0$. From \equ{PRgaussSEC}, one has that $de/dt=0$ ($di/dt=0$) for $e=0$ ($i=0$).

\begin{figure}
\begin{center}
\includegraphics[width=.55\linewidth]{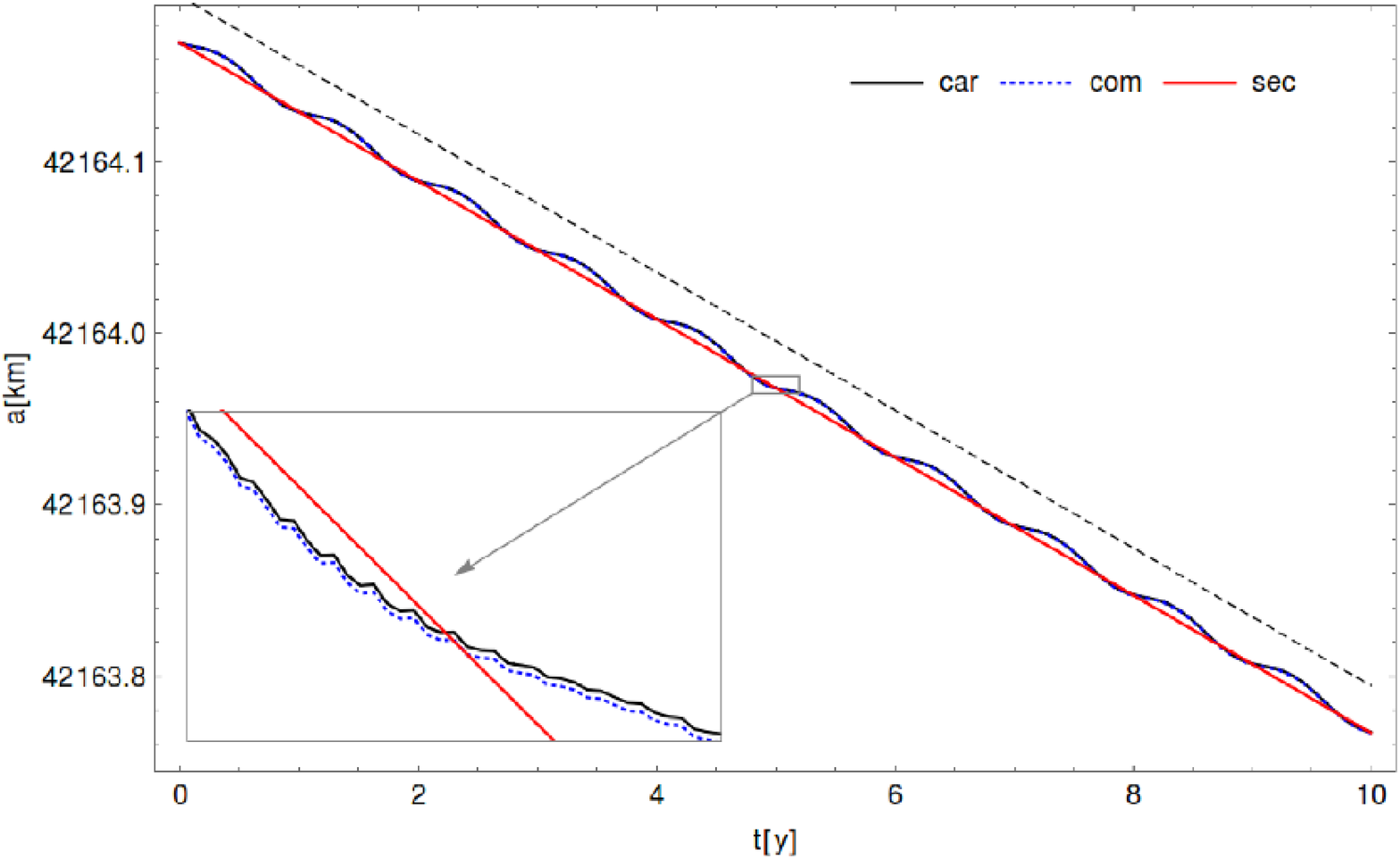}
\end{center}
\caption{Comparison of Cartesian equations including the 2-body problem and PR/SW-drag
(black-thick), complete Gaussian (blue-dotted), and secular (red-thick) model.
Black, dashed: slope of the drift based on the analytical estimate \equ{drift2}.
Lower left: magnification of the central part of the figure.}
\label{toy}
\end{figure}

Equations \equ{PRgaussSEC} lead us to conclude that PR/SW-drag reduces the orbital
energy (and semi-major axis $a$), circularizes the orbit, and may decrease or
increase the inclination in the Sun-Earth system, depending on the orientation
(prograde or retrograde motion) of the orbit.
As a consequence  of the fact that the ratio $n_S/n$ is much smaller
than $a/a_S$, equations \equ{PRgaussSEC} show that the variation of the semimajor axis
is much larger than the variation of the eccentricity and the inclination.
In fact, we notice that $de/dt$,
$di/dt$ are orders of magnitude smaller compared to $da/dt$ due to the common
factor $n_S/n$, and the additional $1/a_S^2$ factor in the right hand sides in
the reduced set of Gaussian equations \equ{PRgaussSEC} (e.g. for a geostationary
orbit in the Sun-Earth system we have $a/a_S^2\propto10^{-8}$, while
$n_S/(na_S^2)\propto10^{-10}$).
This simple remark allows one to obtain as follows the rate of variation of the
semimajor axis. Indeed, in a first approximation we may hold
fixed $e$, $i$, and directly integrate the first of \equ{PRgaussSEC} with respect to time
$t$ to get:
$$
a(t)=a(0)\exp(-C t)
$$
with
\beqno
C=\frac{2\G m_S}{a_S^2}\frac{\beta}{c}\left(1+\frac{\eta}{Q}\right)\left[1+\frac{e_S^2}{2}
-\cos i \cos i_S\left(1-\frac{e^2}{2}+\frac{5e_S^2}{2}\right)\frac{n_S}{n}\right] .
\eeqno
Since $0<C<<1$ we find that, up to first order in $C$, one has
$$
a(t) \simeq a(0)\left(1-Ct\right) \ .
$$
Henceforth, the linear drift rate in semi-major axis caused by PR/SW-drag is given by
\beq{drift2}
\delta a = - a(0) C \ .
\eeq
We evaluate \equ{drift2} for $A/m=1\ [mt^2/kg]$, $Q=1$, $\eta=0$, and find typical drift
rates of the order of $40\ mt/y$ for parameter values of the Sun-Earth system
($a_S=3550[a_{geo}]$, $e_S=0.02$, $i_S=23.45{}^o$, $n_S=1/365[d]$), and initial
conditions $a(0)=a_{geo}\equiv 42\,164.17\ [km]$, $e(0)=0.1$, $i(0)=2{}^o$, and vanishing initial
angles. We compare the outcome of a numerical integration of a model including just the 2-body
problem with $F_{PR/SW}$ drag, with the orbit obtained from the numerical
integration of the system \equ{PRgauss}, \equ{PRgauss2}, and \equ{PRgaussSEC} in Figure~\ref{toy}.
We clearly see that \equ{drift2} well predicts the slope of the drift (dashed, black) of
the numerically obtained solutions.

We remark that on secular time scales ${\mathfrak f}_l$, ${\mathfrak f}_g$, ${\mathfrak f}_h$
vanish. Moreover, if we restrict our analysis to the perturbed two-body problem only,
the total time derivative, given by
\beq{dHdt}
\Phi=
\frac{\partial {\mathfrak H}}{\partial L}\frac{dL}{dt} +
\frac{\partial {\mathfrak H}}{\partial G}\frac{dG}{dt} +
\frac{\partial {\mathfrak H}}{\partial H}\frac{dH}{dt} +
\frac{\partial {\mathfrak H}}{\partial l}\frac{dl}{dt} +
\frac{\partial {\mathfrak H}}{\partial g}\frac{dg}{dt} +
\frac{\partial {\mathfrak H}}{\partial h}\frac{dh}{dt} +
\frac{\partial {\mathfrak H}}{\partial t} \ ,
\eeq
will reduce to the simple form:
\beq{dHdtsec}
\Phi = -\frac{\mu}{2a}C \ .
\eeq
We remark that $\Phi$ is zero just for ${\vec F}_{PR/SW}=0$, while the variation
of $\Phi$ in the dissipative problem allows one to
quantify the overall effect of combined solar radiation pressure and
Poynting-Robertson drag in weakly dissipative, non-integrable dynamical
systems (\cite{CL2012, LC2013}).

\subsection{Equilibria and linear stability analysis}
\label{subsec:equilibria}

\begin{figure}
\begin{center}
\includegraphics[width=.55\linewidth]{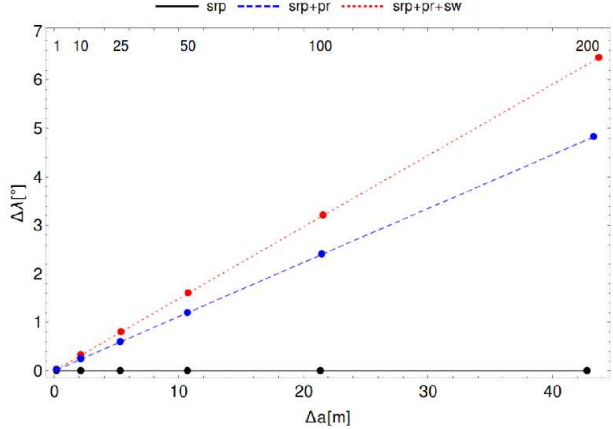}
\end{center}
\caption{Location of the equilibria in the $(\Delta a,\Delta\lambda)$-plane
for $A/m=1$, 10, 25, 50, 100, 200 (small ticks): solar radiation
pressure alone (black, thick), with the additional Poynting-Robertson effect
(blue, dashed), all together with solar wind drag (red, dotted), i.e. $\eta=1/3$,
$Q=1$. Additional Kepler elements have been set to $e=0$, $i=0{}^o$, $\omega=0{}^o$,
$\Omega=0{}^o$, respectively.}
\label{equ}
\end{figure}

To highlight the role of solar radiation pressure and PR/SW-drag on
the location of the equilibrium of the geostationary orbit we make use of a
simplified system of equations that qualitatively describes the motion of the
SDO close to the geostationary orbit, i.e. close to the 1:1 resonance between the
orbital period of the SDO and the rotational period of the Earth.
First, we omit the gravitational effect of the Sun and
the Moon on the motion of the SDO, and just take the gravitational field of the
Earth up to degree and order $2$, solar radiation pressure, and the combined
PR/SW-drag terms of \equ{eq1} into account. Next, we derive
the components $da/dt$, $dM/dt$ of \equ{gauss} that are needed to get the
components $dL/dt$, $dl/dt$ of \equ{HAM} for both, the gravitational field of
the Earth, $V_E$, and the solar radiation pressure term $V_{SRP}$. Let
$\tilde\omega=\omega+\Omega$, and again let $\theta$ be the Greenwich Meridian angle. We
introduce the resonant argument for the stationary orbit as
\beqno
\lambda=M-\theta+\tilde\omega\ .
\eeqno
Next we insert $\lambda$ into the expression for $dL/dt$, $dl/dt$ in \equ{HAM} and average over the rotation period $\theta$.
Let $\mathfrak H_{res}$, $\mathfrak f_{L,res}$, $\mathfrak f_{l,res}$ denote the
resulting averaged terms. The equilibrium that defines the stationary orbit in this
simplified resonant system is provided by the set of equations:
\beqa{ressys}
\frac{dL}{dt} = - \frac{d \mathfrak H_{res}}{d \lambda} + {\mathfrak f}_{L,res} = 0 \ , \quad
\frac{d\lambda}{dt} = \frac{d \mathfrak H_{res}}{d L} + {\mathfrak f}_{l,res} = 0   \ ,
\eeqa
where the last relation holds up to a constant and where we assumed that
$d\tilde\omega/dt$ is zero.
We remark that in equations \equ{ressys} the terms in $\mathfrak H_{res}$ are
stemming from $V_E$, $V_{SRP}$, while the terms in $\mathfrak f_{L,res}$,
$\mathfrak f_{l,res}$ originate from $\vec F_{PR/SW}$ in \equ{eq1}. Since \equ{ressys}
still depends on all Keplerian elements, we fix all but $a$, $\lambda$, for
which we solve for given system parameters. We present our results in the
($\Delta a, \Delta \lambda)$ -plane, where $\Delta a=a_c-a_*$, $\Delta
\lambda=\lambda_c-\lambda_*$ with $a_c$, $\lambda_c$ being the solution of the
classical problem (i.e., without PR/SW-drag), and $a_*$, $\lambda_*$ are the
values that solve \equ{ressys} including the PR/SW-drag. Our results are
summarized in Figure~\ref{equ}. We clearly see that solar radiation pressure
and the drag terms together shift the location of the geostationary equilibrium up to $50\ mt$ in semi-major
axis, and up to $6{}^o$ in orbital longitude. If we neglect the drag terms, but still
take into account the solar radiation pressure terms, the shift in semi-major
axis persists. Therefore, we conclude, that the shift in semi-major axis is mainly caused
by the effect of solar radiation pressure alone, while the shift in orbital longitude
is necessary to balance the additional effect of the drag terms.
Figure~\ref{equ} also shows that for the parameter values used in the figure,
the role of solar wind is not negligible for high area-to-mass ratios.

\begin{figure}
\begin{center}
\includegraphics[width=.55\linewidth]{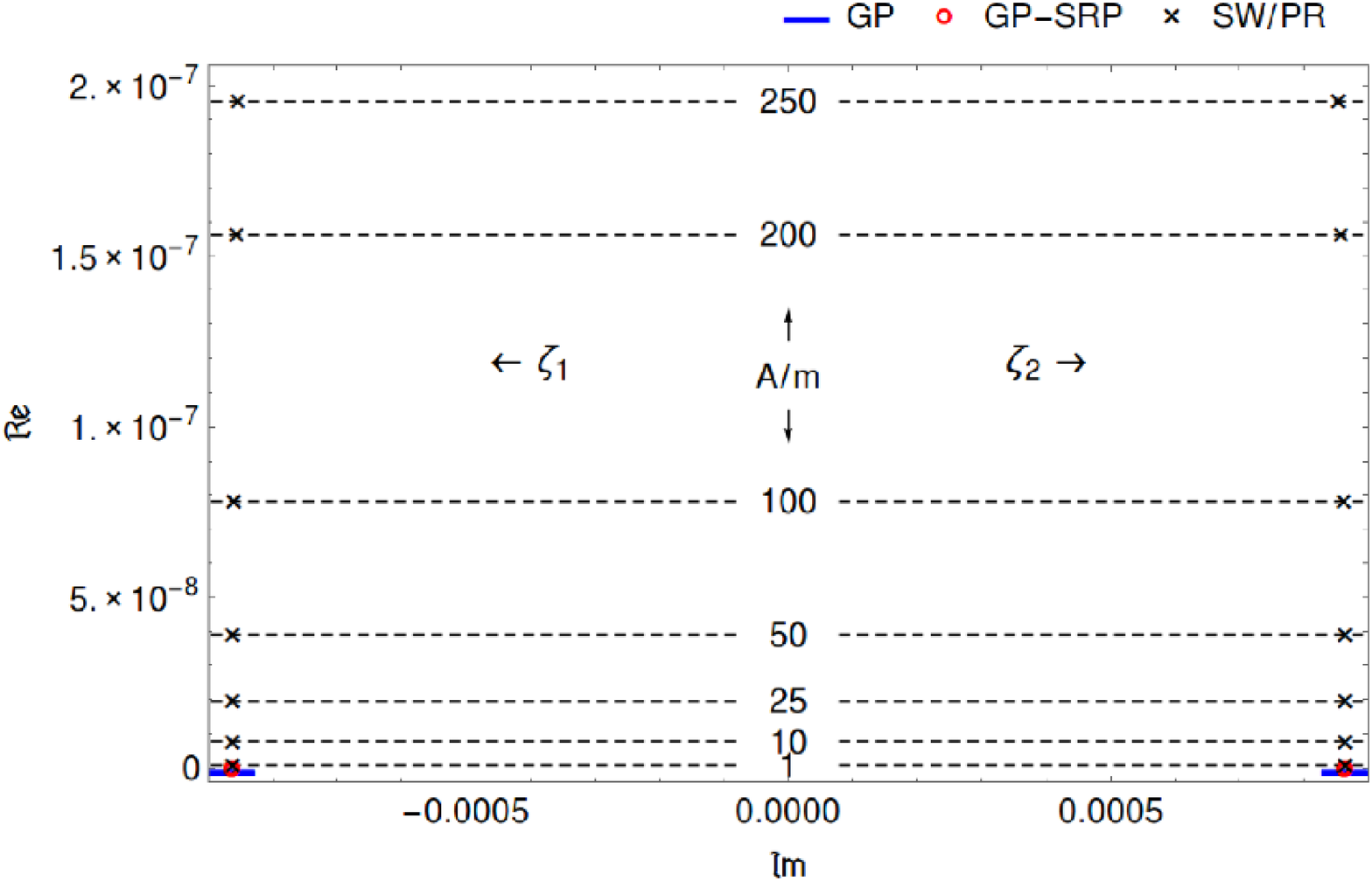}
\end{center}
\caption{Eigenvalues, in the $({\mathfrak Im}, {\mathfrak Re})$-plane, of the linearized
system of \equ{ressys} expanded around the stationary point for different $A/m$ (indicated
by small numbers inside the plot), for the same circular orbit as in Figure~\ref{equ}.}
\label{sta}
\end{figure}

Next, we linearize the left hand sides of \equ{ressys} around the stationary
solution, and calculate the eigenvalues $(\zeta_1, \zeta_2)$ of the
linearized system for the same parameters and initial conditions as in the
study of the equilibria. Our results are provided in Figure~\ref{sta}: we
clearly see, that the combined PR/SW-drag effect introduces a positive real
part to $(\zeta_1, \zeta_2)$ growing with increasing $A/m$ ratio. For
comparison, we also derive the eigenvalues on the basis of $V_E$, as well as of $V_E$ and
$V_{SRP}$ in \equ{ressys}. In both cases, the eigenvalues stay complex,
yielding the elliptic character of the equilibrium for circular orbits.

To conclude, let us mention that in the following we will refer to a GEO 1:1 resonance, whenever
the following relation holds:
$$
\dot M-\dot\theta=0\ ,
$$
where $\dot M$ corresponds to the mean motion of the SDO, while $\dot\theta$ denotes the angular speed
of rotation of the Earth. Such resonance corresponds to a geostationary orbit located on the equatorial
plane at a distance of about 42\,164.17 $km$ from the center of the Earth.

\section{Temporary trapping or escape from the resonance}\label{num}

In this section we perform several numerical experiments to study the behavior
of the orbits close to the GEO 1:1 resonance. As it is well known, within the
conservative framework the GEO 1:1 resonance shows a pendulum-like phase
portrait. Indeed, the exact 1:1 resonance corresponds to the equilibrium point, which
is surrounded by a librational island, whose border is delimited by a chaotic
separatrix (\cite{CG2014a}). When the dissipation is switched on, the orbits
might collapse on the equilibrium, can be temporarily trapped in a resonant regime,
or rather escape from the resonance (\cite{Neis}). Although the PR-SW drag is rather
weak, we observe that its effect is not negligible on a proper time scale, as shown in
the Sections~\ref{sec:drift} and \ref{sec:behavior}.

Let us premise some information about the parameters and data used in the
forthcoming numerical simulations: the astronomical constants $\G m_E$, $\G m_M$,
$G m_S$ are taken from \cite{AstCon}, the gravity field of the Earth
(tide-free gravity field EGM2008) are obtained from \cite{EGM2008, PavEtAl2012}. The
ranges for various parameters related to PR/SW-drag are derived on
the basis of values found in the literature (see \cite{BLS1979,
Gus1994, Koc2006}). We remark that for typical optical properties and
densities, that are consistent with observations,
$\beta\simeq0.2/s$ (with the radius $s$ of a spherical particle
given in $[\mu mt]$, see, e.g. \cite{BeaFer1994}), or
$\beta\simeq7.6\times10^{-4}A/m$ (with $A/m$ given in $[mt^2/kg]$,
see, e.g. \cite{Koc2006}) may be used.

In all numerical simulations, the initial Epoch is J2000 (January 1, 2000, 12:00 GMT).
We provide the numerical results in terms of osculating orbital elements.
The transformation between inertial Cartesian coordinates and osculating Kepler
elements is summarized in Appendix A.

%
%

\subsection{Drift motion (outside resonances)}\label{sec:drift}

\begin{figure}
\begin{center}
\includegraphics[width=.45\linewidth]{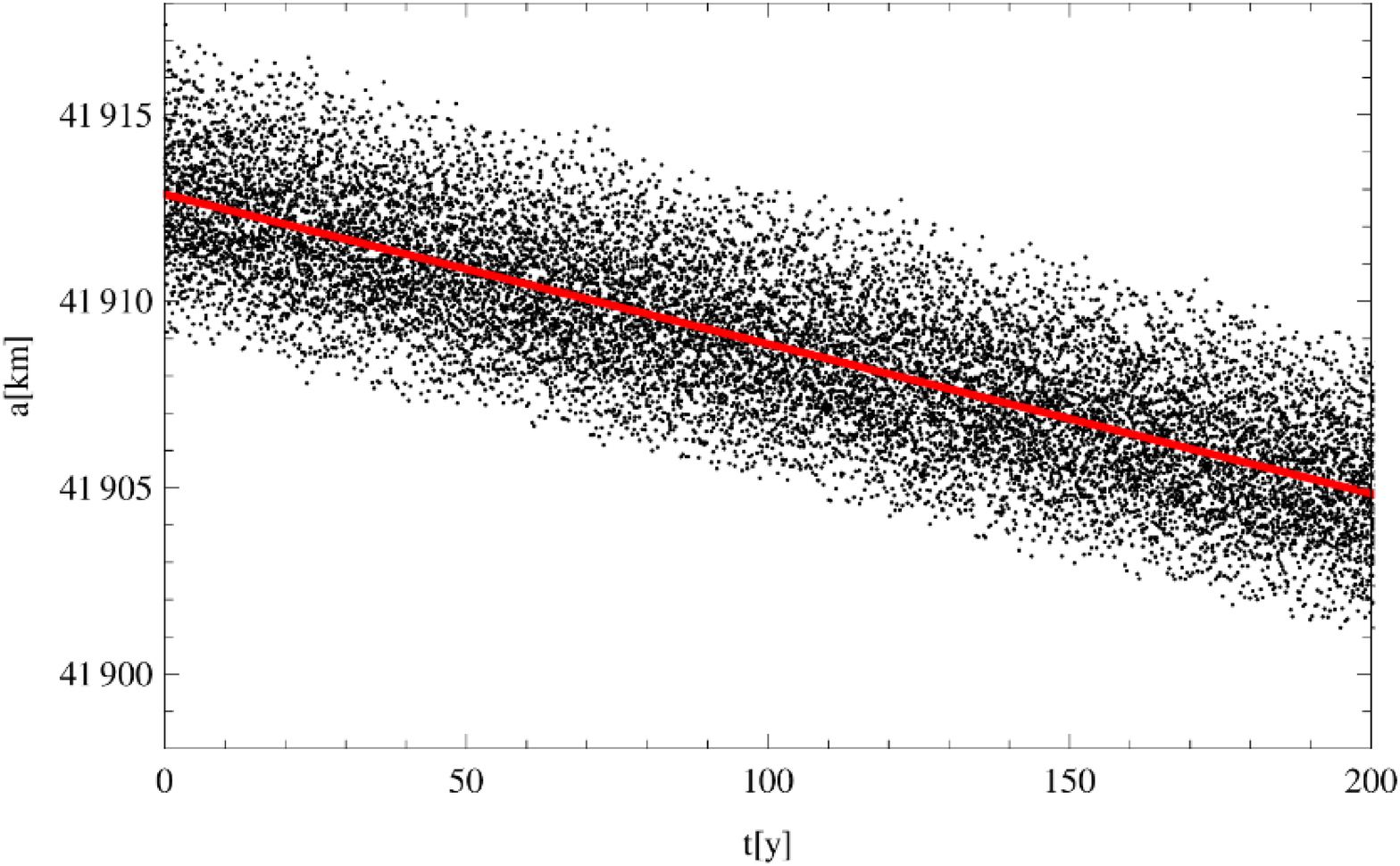}
\includegraphics[width=.45\linewidth]{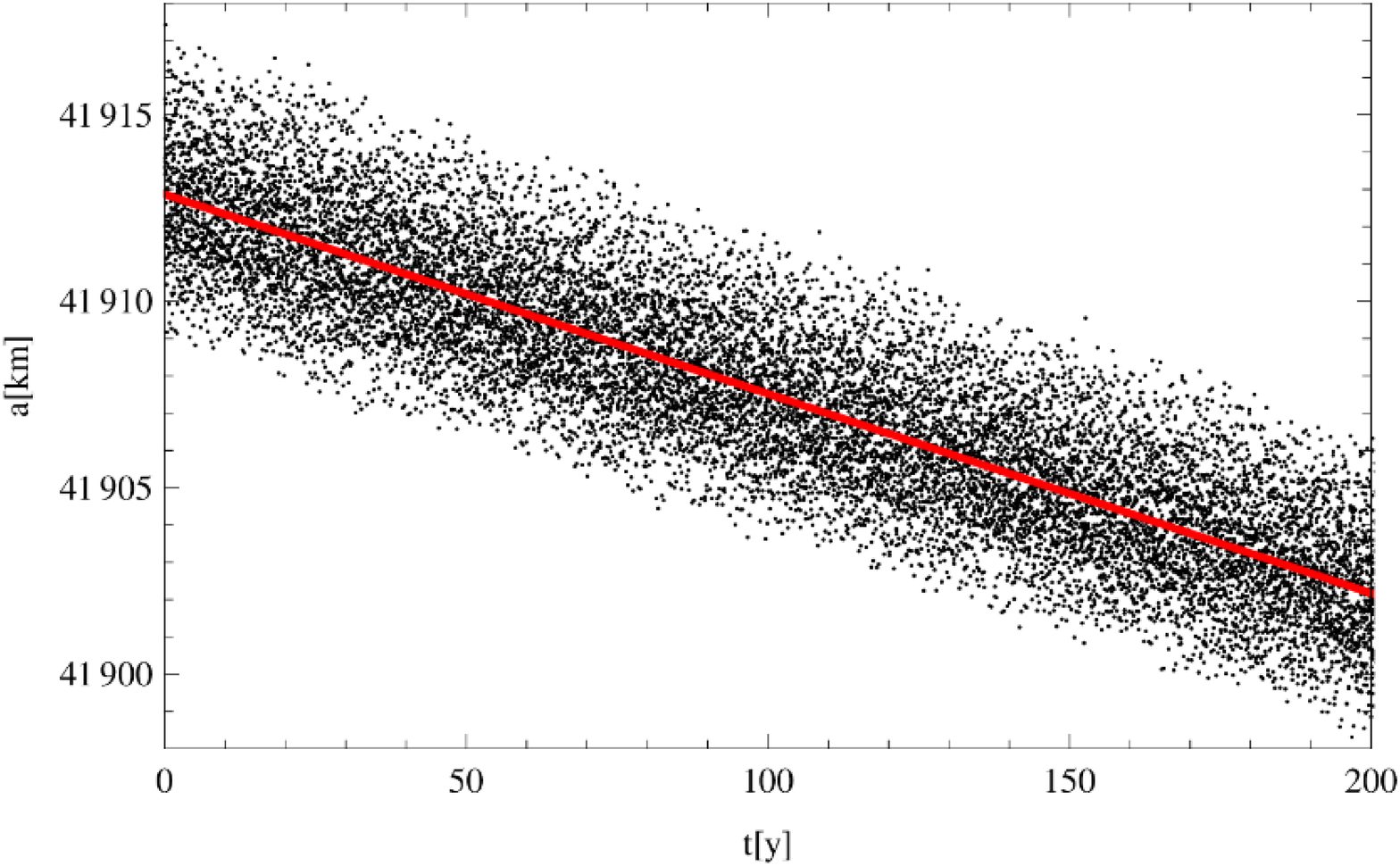}
\end{center}
\caption{Typical drift motion in semi-major axis $a$ outside a resonant regime.
Parameters and osculating initial conditions are: $A/m=1\, [mt^2/kg]$, $a(0)=41914.1696\
[km]$, $e(0)=0.2$, $i(0)=0{}^o$,
$\omega(0)=0{}^o$, $\Omega(0)=0{}^o$, $M(0)=0{}^o$. Left: $\eta=0$
(without solar wind). Right:  $\eta=1/3$, $Q=1$ (with solar wind). The
slope of the drift rate calculated on the
basis of \equ{drift2} is shown in red-thick.}
\label{drift}
\end{figure}

Outside a resonant regime, the dynamical behavior leads to a drift of the
semimajor axis.  This is well explained by the following example of drift
motion, which is shown in Figure~\ref{drift}, obtained by a direct numerical
integration of \equ{eq1} (including all effects).   The drift in semimajor axis
is conveniently described by the secular theory developed in Section~\ref{qua}, i.e.
formula \equ{drift2}.

\begin{figure}
\begin{center}
\includegraphics[width=.55\linewidth]{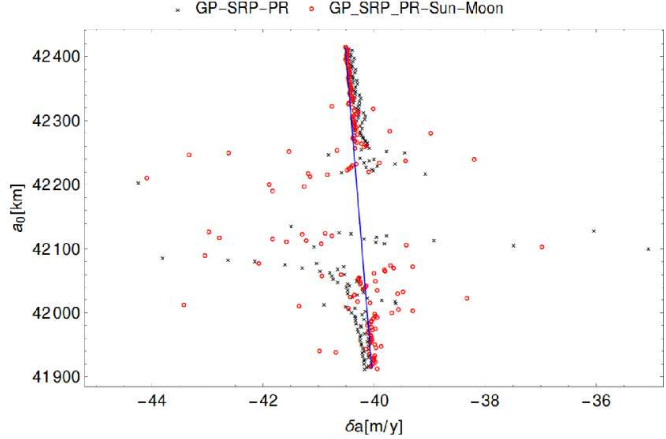}
\end{center}
\caption{Drift rates close to the GEO regime of motion. Numerically obtained
drift rates on the basis of \equ{eq1} for osculating initial conditions within
$a(0)=a_0$ and $41914.1696\ [km]\leq a_0\leq42414.1696\ [km]$, and same parameters
and additional initial conditions as in Figure~\ref{drift}. Black-cross:
without the additional effect of the Sun and the Moon. Red-circle: full
model. Blue-thick line: drift rates calculated on the basis of \equ{drift2}.}
\label{driftrates}
\end{figure}

We provide a survey of the drift rates in
Figure~\ref{driftrates}: while in the rotational regime of motion of the
resonant angle $\lambda$, the drift due to the combined PR/SW-effect
is essentially described by a constant linear drift, the drift rates are spread,
when getting close to the exact resonance. Figure~\ref{driftrates} also shows
that the role of Moon and Sun becomes relevant outside the resonance, as
already noticed, e.g., in \cite{CG2014b}, \cite{Rosengren}.

\subsection{Behavior in the neighborhood of the GEO 1:1 resonance}\label{sec:behavior}

Depending on parameters and initial conditions, the numerical study unveils a
rich dynamical behavior in the vicinity of the 1:1 resonance, much more complex
than the linear drift described above, consisting of {\it trapped motions} in
primary or higher order resonances, {\it resonance captures}, {\it escapes from
resonance} and {\it jumps}. Although we know that such behavior is typical of dissipative systems,
we aim to show that PR-SW drag is not negligible on reasonably long time
scales.  For instance, Figure~\ref{fig:Am1}, upper plots, will show the stabilizing
effect of the resonance condition between the orbital period of the satellite
and the rotational period of the Earth on the long-term motion of the SDO
(temporarily trapped motion into primary resonance). We clearly see that
initial conditions starting in the librational regime of motion of the resonant
angle $\lambda$ stay close to their initial orbital semi-major axis $a(0)$
on long time scales.

In order to point out the role of the PR--drag effect close to the 1:1
resonance and to depict numerically all the above mentioned phenomena, we consider
two sample objects having $A/m=1\, [mt^2/kg]$ and $A/m=15\, [mt^2/kg]$,
respectively. For spherical bodies having the typical density $\rho=2.2 \,
[g/cm^3]$, these values of the area-to-mass ratio correspond to debris of the
order of sub-millimeters in diameter.

To avoid complex effects induced by the lunisolar secular resonances (see for
example \cite{Hughes(1980), Cellettietal(2016), Daquinetal(2016)}), we focus on
a region of the space of orbital elements characterized by small inclinations
and not very large eccentricities, where instead lunisolar secular resonances
might strongly influence the dynamics.  As a matter of fact, in computations,
we took the initial inclination $i(0)=10^o$ and the initial eccentricity
$e(0)=0.2$. Since the eccentricity is not zero, for large values of the
area-to-mass ratio, like $A/m=15\, [mt^2/kg]$, some secondary resonances
between the geostationary libration angle $\lambda$ and the Sun's longitude,
denoted hereafter by $\lambda_{Sun}$, appear as effect of the solar radiation
pressure (\cite{Valketal2009, Lem2009}). We remark that the term \sl secondary
\rm is commonly used to describes resonances within a librational regime, while
here - in agreement with the terminology adopted in \cite{Valketal2009} - we
label \sl secondary \rm those resonances which are due to an interaction between the
geostationary libration angle and the longitude of the Sun.

\begin{figure}
\centering
\vglue0.1cm
\hglue0.1cm
\includegraphics[width=.45\linewidth]{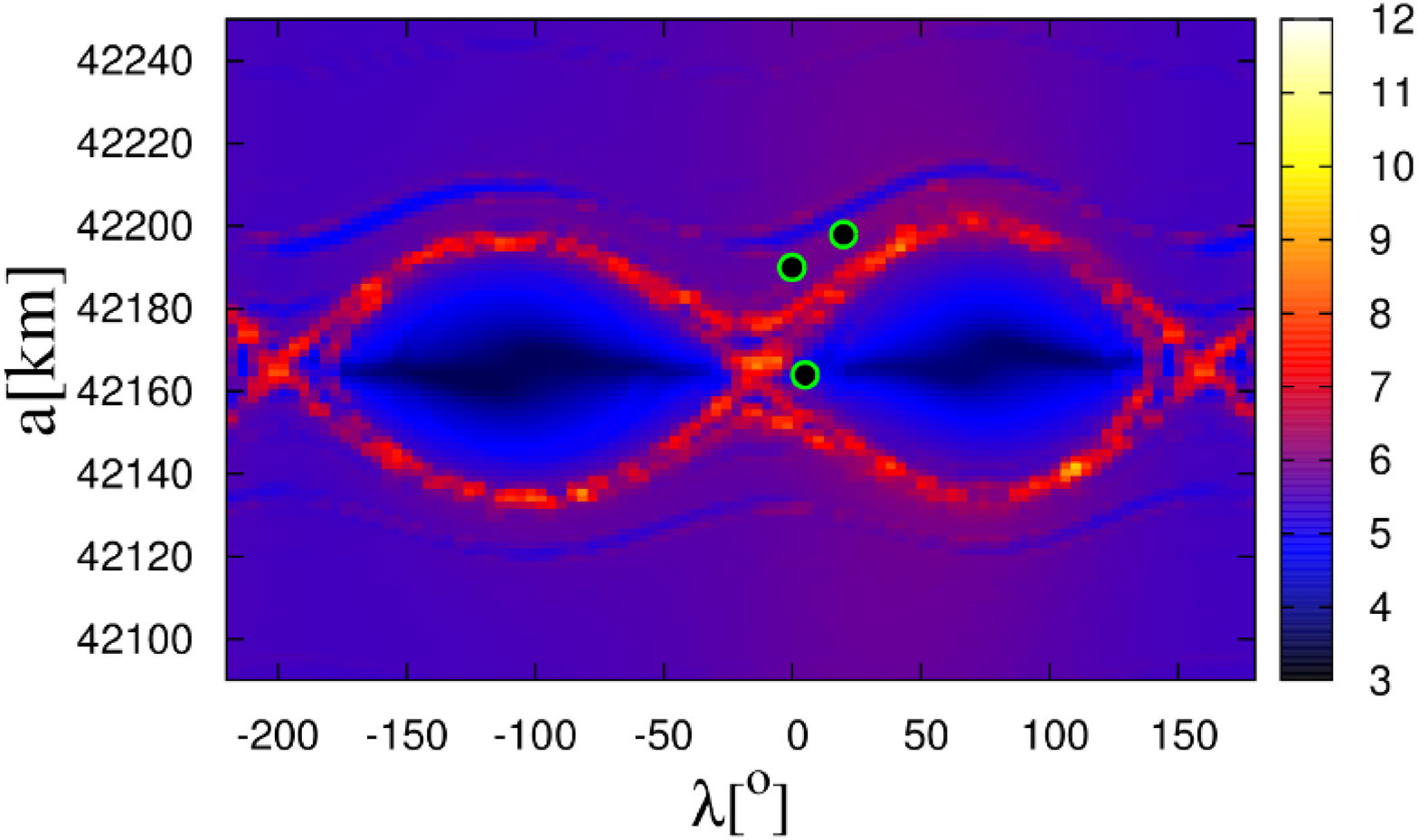}
\includegraphics[width=.45\linewidth]{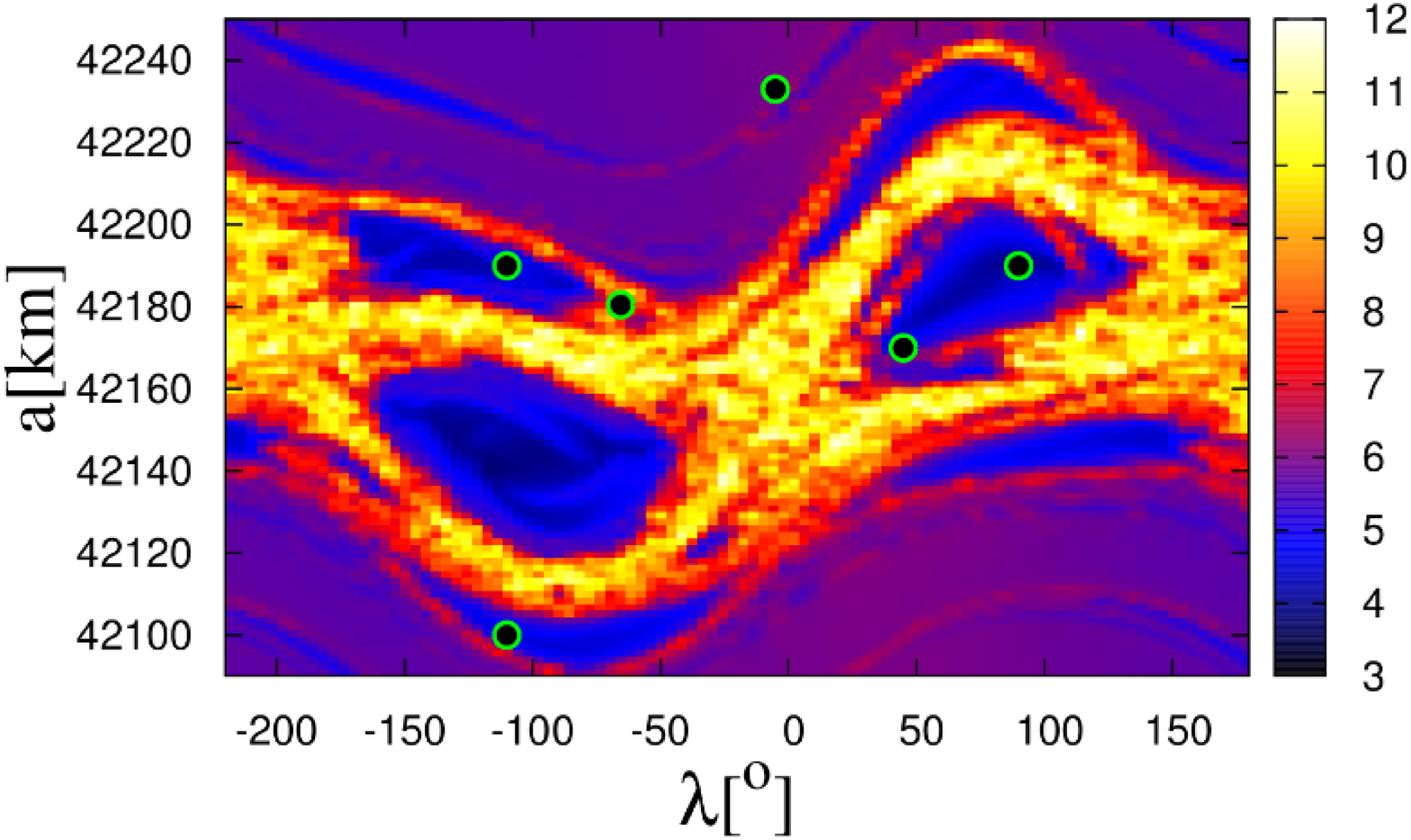}
\vglue0.5cm
\caption{FLI map (using Cartesian equations) for the GEO 1:1 resonance, under
the effect of the following perturbations: solar radiation pressure, Earth's
harmonics up to $n=m=3$, Sun and Moon (conservative case). The osculating
initial conditions are: $e(0)=0.2$, $i(0)=10^o$, $\omega(0)=10^o$,
$\Omega(0)=20^o$. Left: $A/m=1\, [mt^2/kg]$; Right: $A/m=15\, [mt^2/kg]$. The
green--black circles represent the orbits analyzed in Figures~\ref{fig:Am1},
\ref{fig:Am15}, \ref{fig:Am15_secondary_resonances},
\ref{fli:Am15_chaotic_region} and \ref{fli:Am15_outside}, in the framework of
the full model which includes also the PR/SW--drag effect.}
\label{fig:FLI_Am1_Am15}
\end{figure}

\begin{figure}
\centering
\vglue0.1cm
\hglue0.1cm
\includegraphics[width=.45\linewidth]{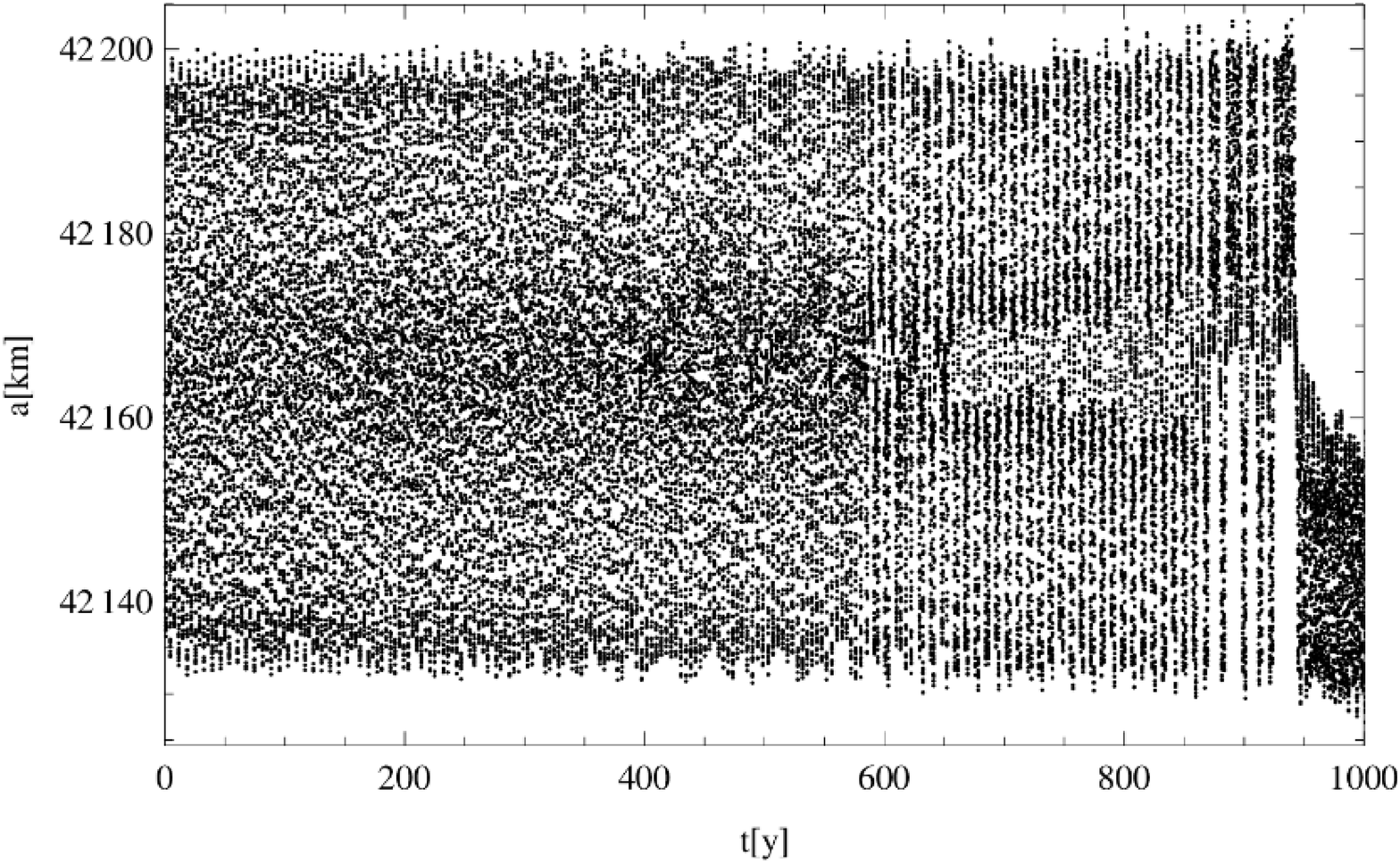}
\includegraphics[width=.45\linewidth]{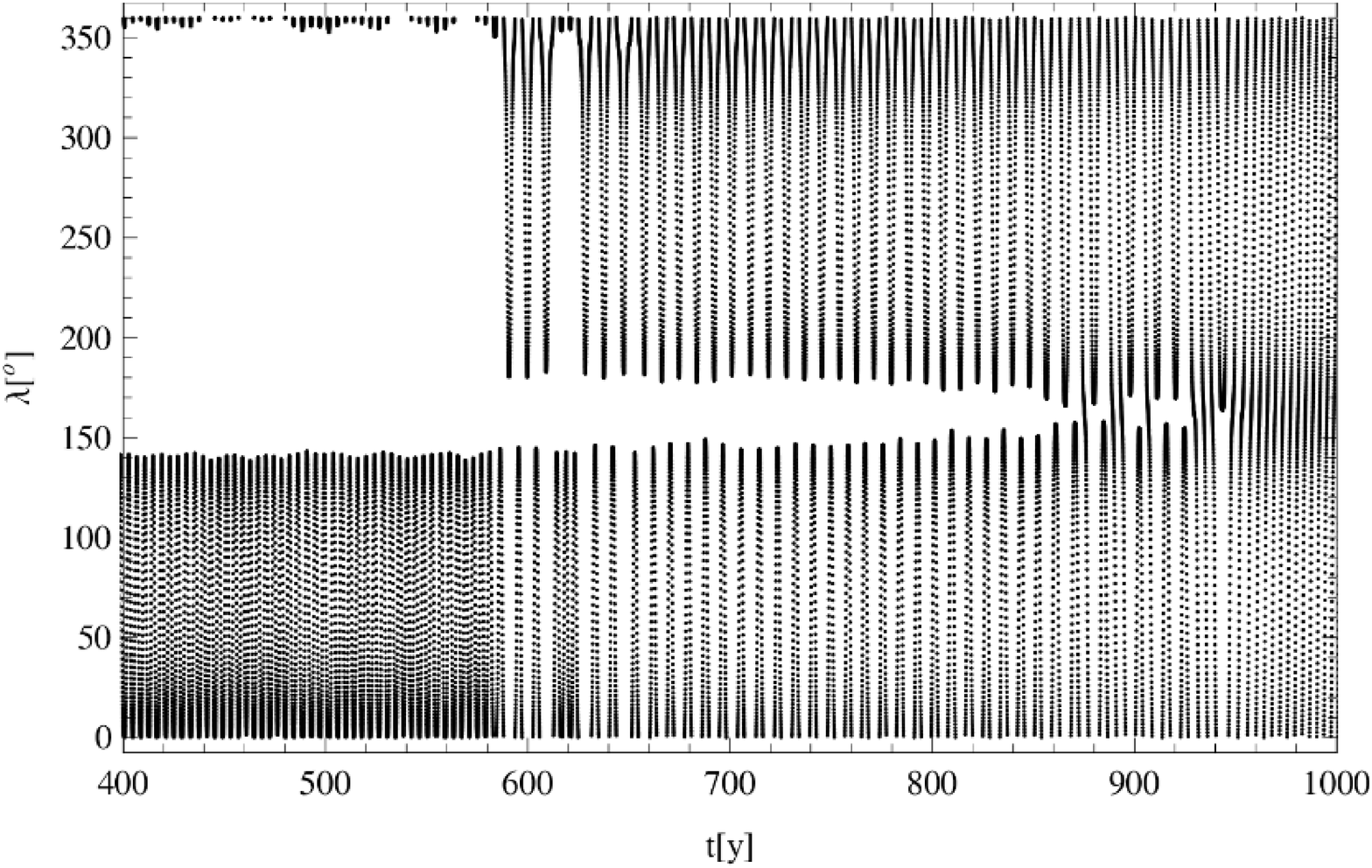}\\
\includegraphics[width=.45\linewidth]{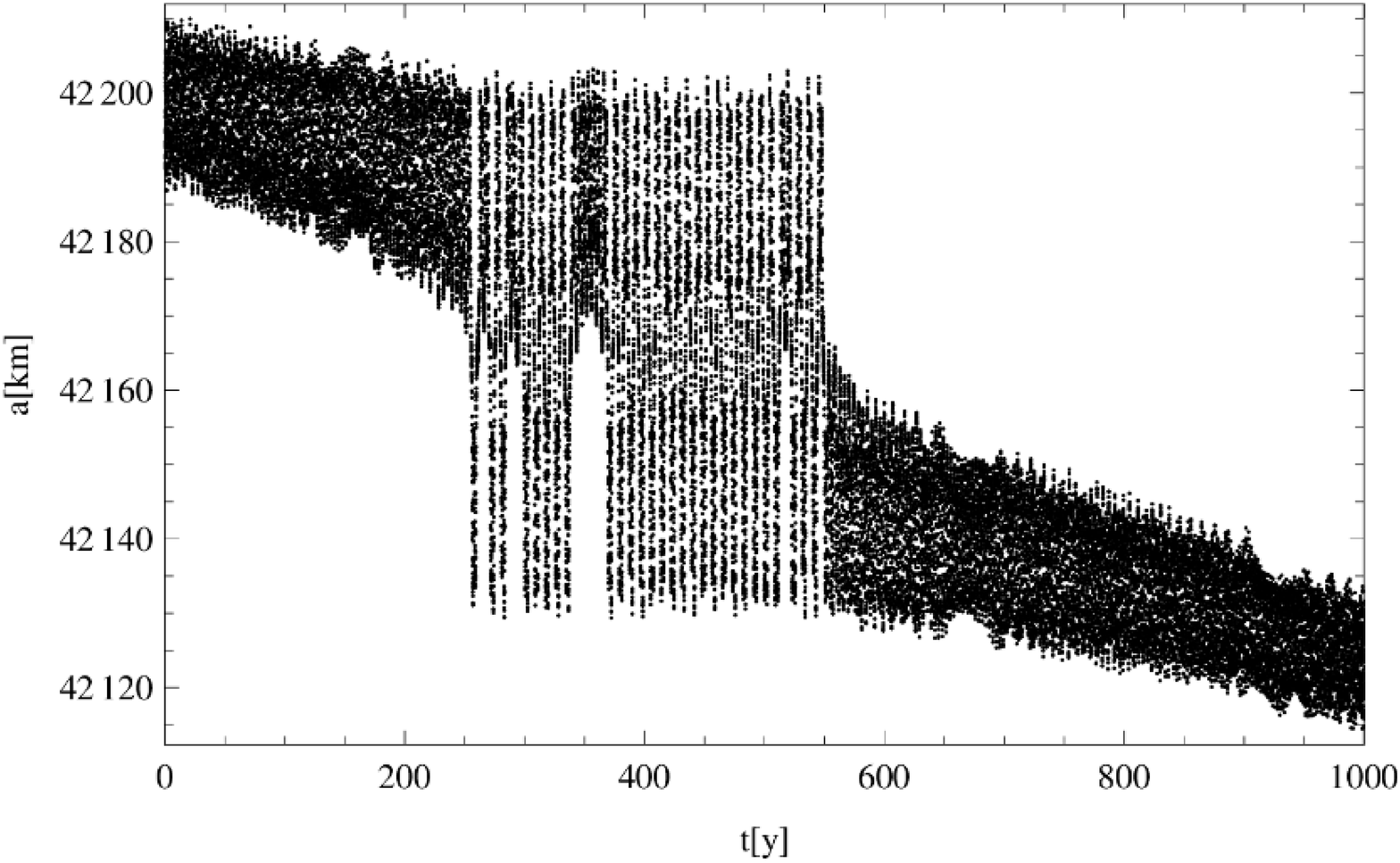}
\includegraphics[width=.45\linewidth]{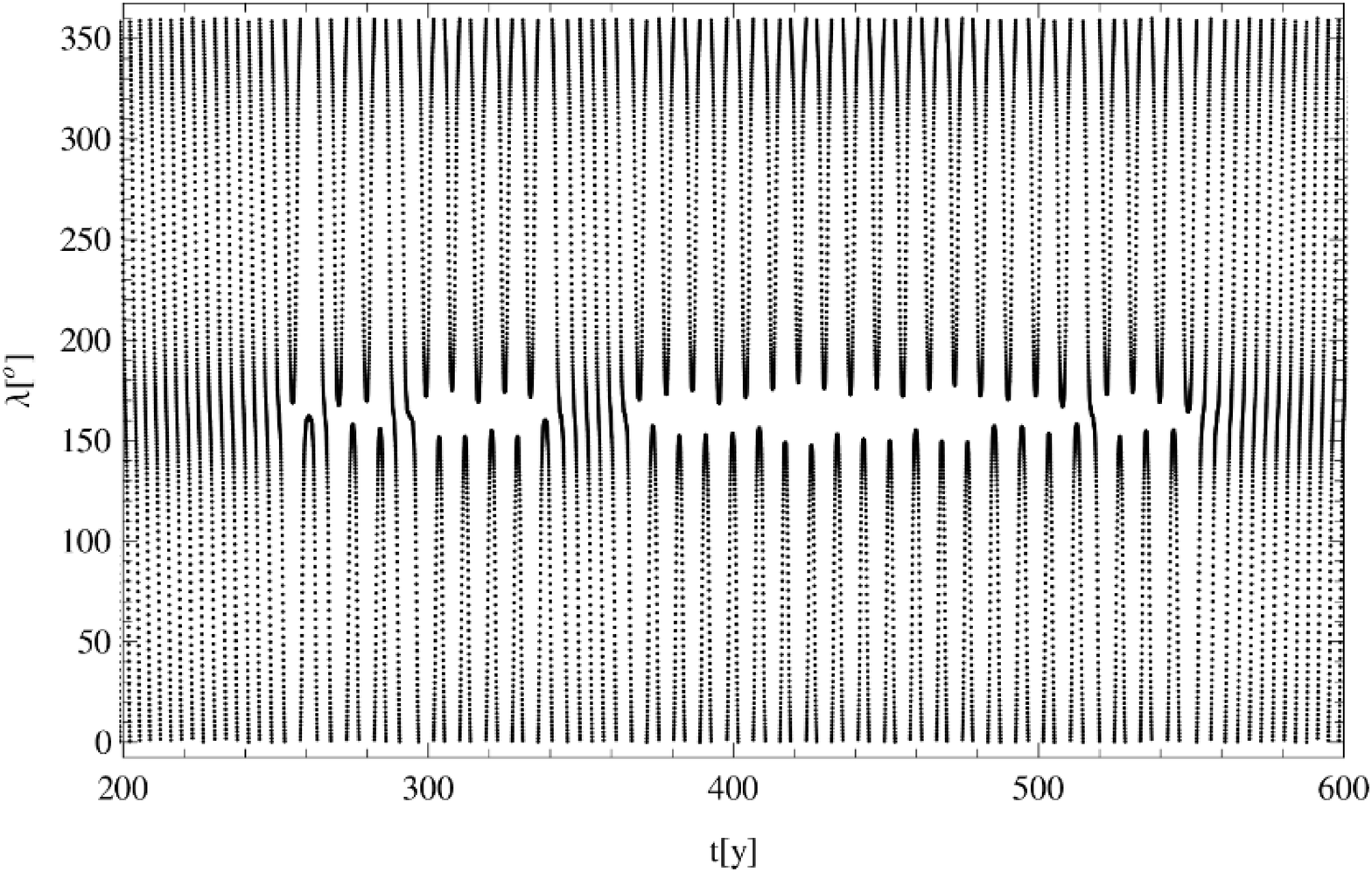}\\
\includegraphics[width=.45\linewidth]{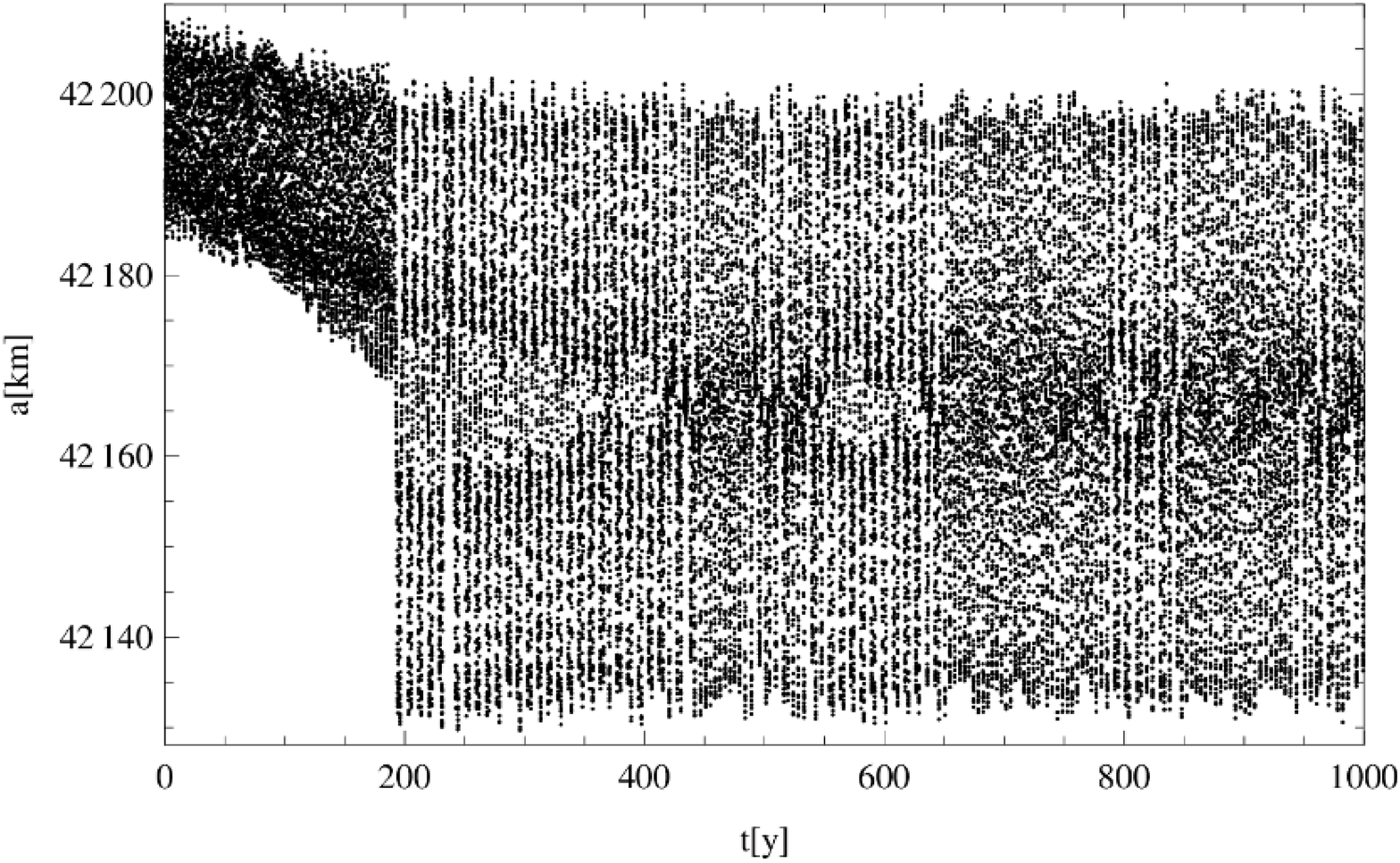}
\includegraphics[width=.45\linewidth]{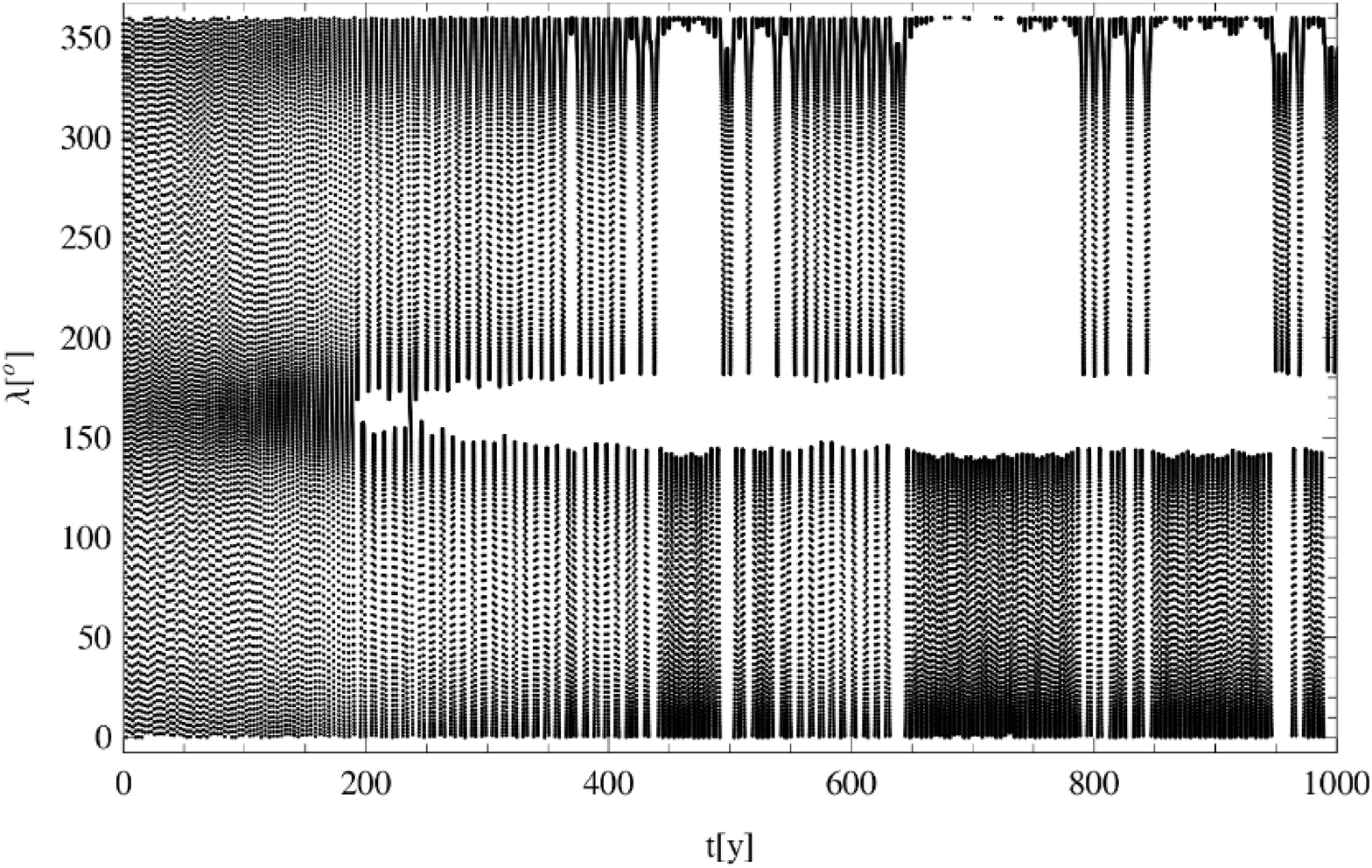}
\vglue0.5cm
\caption{Variation of the semimajor axis $a$ and resonant angle $\lambda$ for three orbits
located in the neighborhood of the 1:1 resonance. Parameters and osculating initial conditions
are $A/m=1\ [mt^2/kg]$, $\eta=0$, $i(0)=10^o$, $e=0.2$, $\omega(0)=10^o$, $\Omega(0)=20^o$ and:
$a(0)=42164\, km$, $\lambda(0)=5^o$ (or $M(0)=-25^o$) (top panels);   $a(0)=42198\, km$,
$\lambda(0)=20^o$ (or $M(0)=-10^o$) (middle panels); $a(0)=42190\, km$, $\lambda(0)=0^o$
(or $M(0)=-30^o$) (bottom panels). Compare with the left panel of Figure~\ref{fig:FLI_Am1_Am15}.}
\label{fig:Am1}
\end{figure}

To highlight the influence of the PR/SW drag, it is important to discuss first
the dynamical effect induced by the other perturbations (namely, by the
conservative part), in particular the solar radiation pressure.
Figure~\ref{fig:FLI_Am1_Am15} computes two FLI plots\footnote{Fast Lyapunov Indicator were introduced in
\cite{FLIref} as a measure of the regular and chaotic behavior of a dynamical system.
Roughly speaking, they are defined as the Lyapunov exponents at finite times. We refer the reader
to \cite{FLIref} for the definition of the FLI and its properties.}
for the Cartesian model
described in Section~\ref{subsec:cartesian}, which includes all perturbations
but PR/SW--drag effect, when the area to mass ratio parameter has the values
$A/m=1\, [mt^2/kg]$ (left panel) and $A/m=15\, [mt^2/kg]$ (right panel). The
(osculating) initial conditions are $e(0) = 0.2$, $i(0) = 10^o$,
$\omega(0)=10^o$, $\Omega(0) = 20^o$. The color scale provides a measure of the FLI,
which gives an indication of the regular or chaotic dynamics: small values
(i.e., dark colors) correspond to regular motions, while larger values
(i.e., red to yellow colors) denote chaotic regions.
The phase plane $\lambda$--$a$ is very
similar to that of a pendulum for  $A/m=1\, [mt^2/kg]$ (left panel of
Figure~\ref{fig:FLI_Am1_Am15}). The semimajor axis $a$ and the resonant angle
$\lambda$ librate or circulate; the red-yellow curves divide the phase space in
regions corresponding to libration or circulation.

\begin{figure}
\centering
\vglue0.1cm
\hglue0.1cm
\includegraphics[width=.45\linewidth]{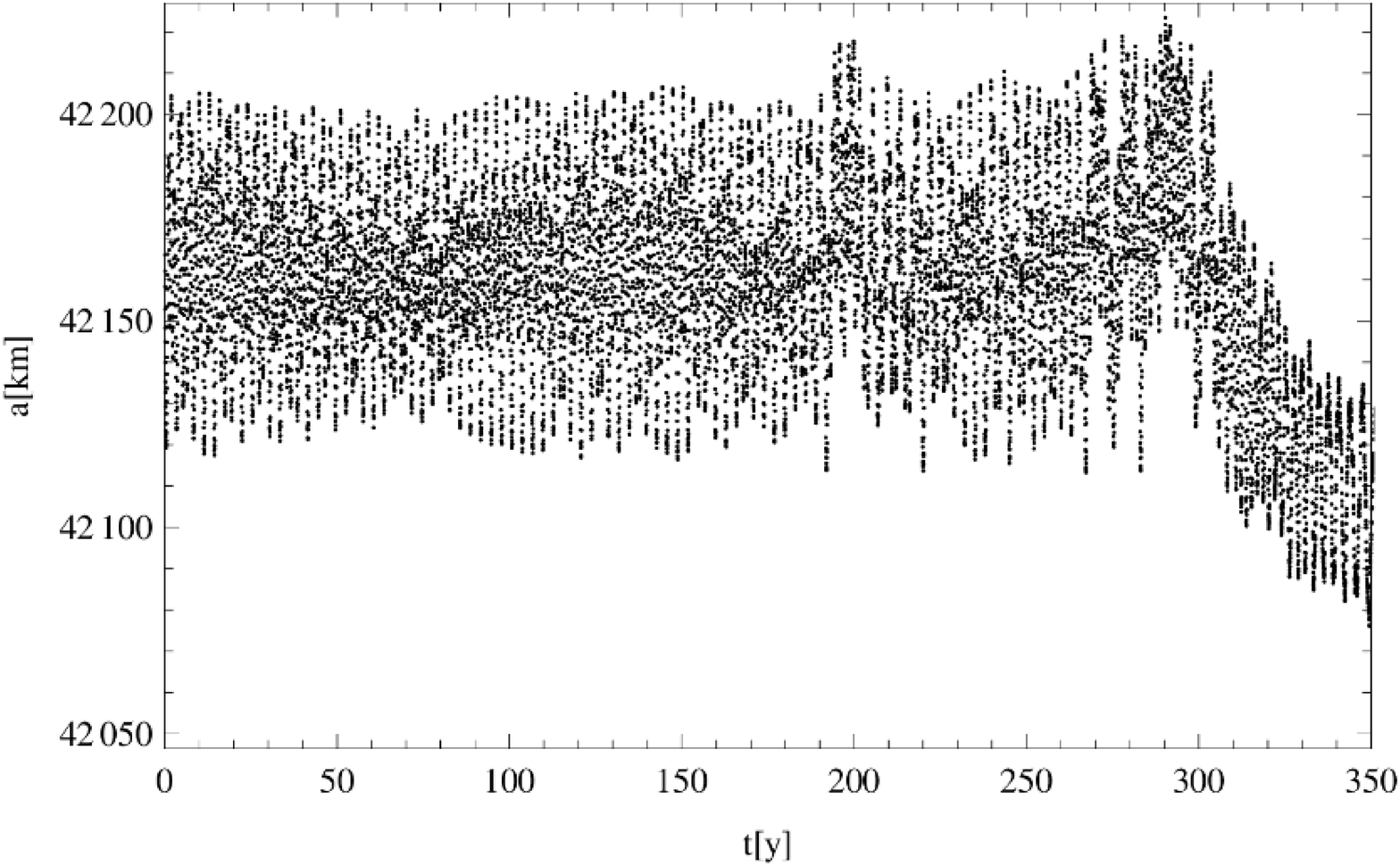}
\includegraphics[width=.45\linewidth]{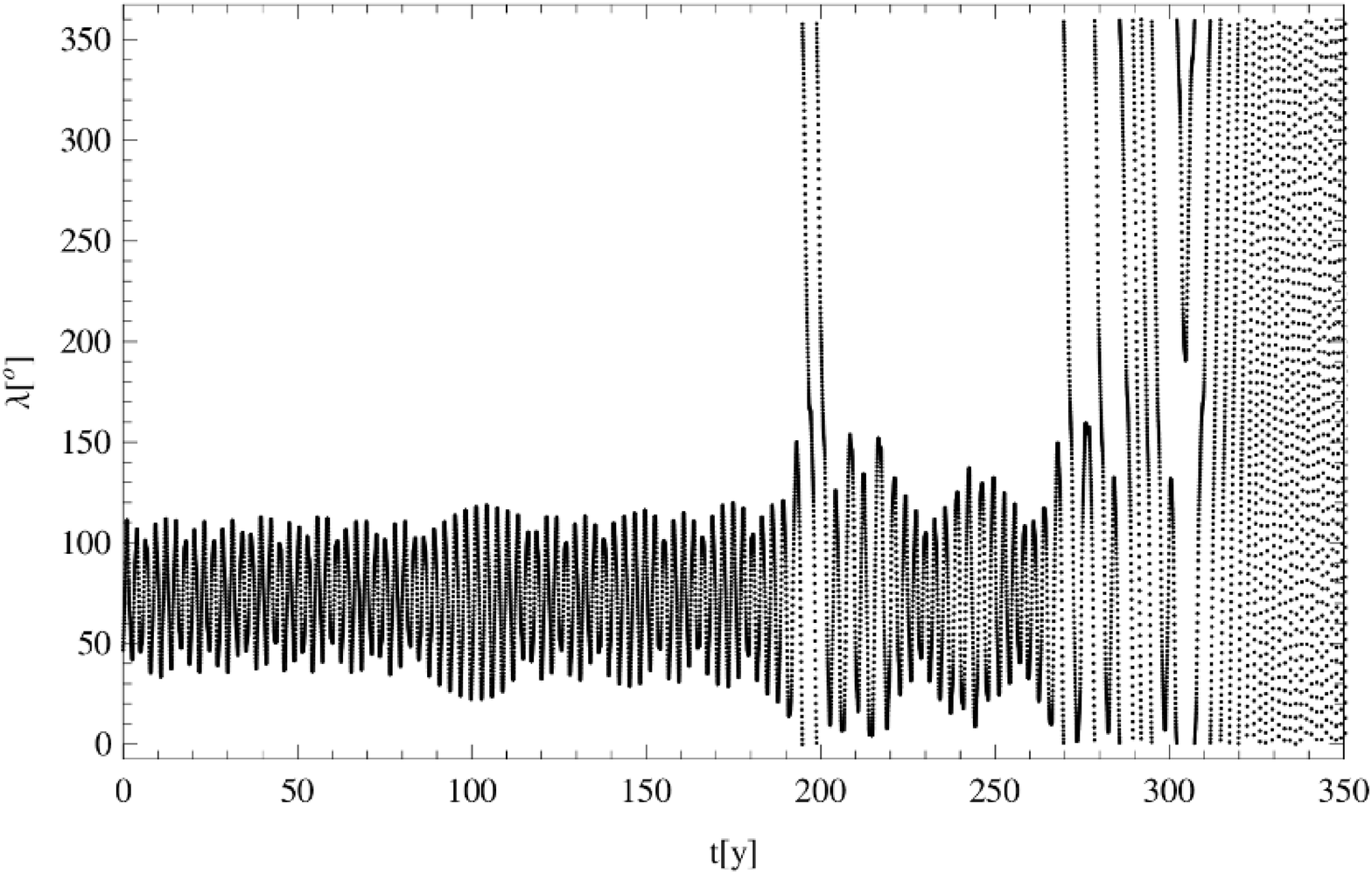}\\
\includegraphics[width=.45\linewidth]{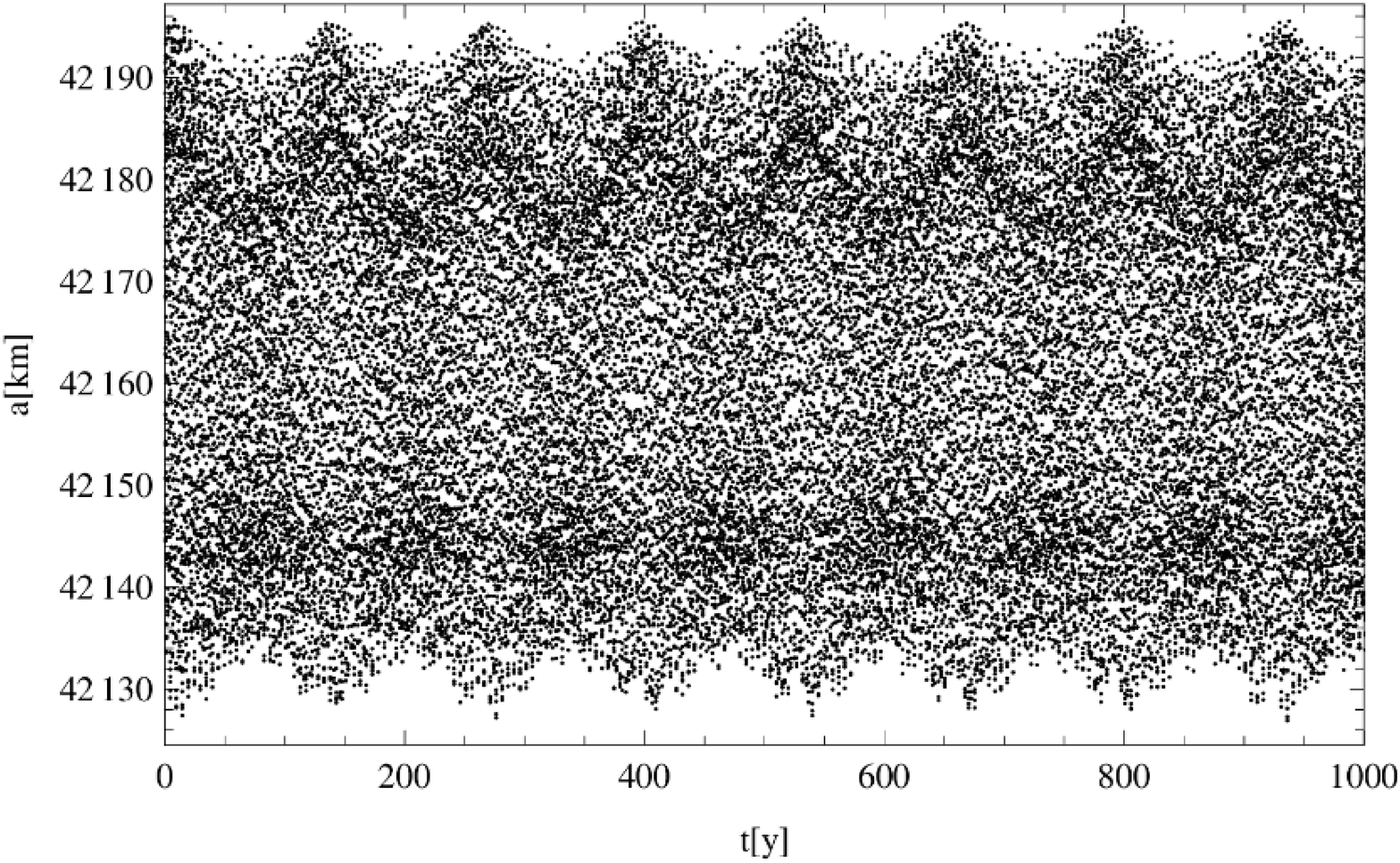}
\includegraphics[width=.45\linewidth]{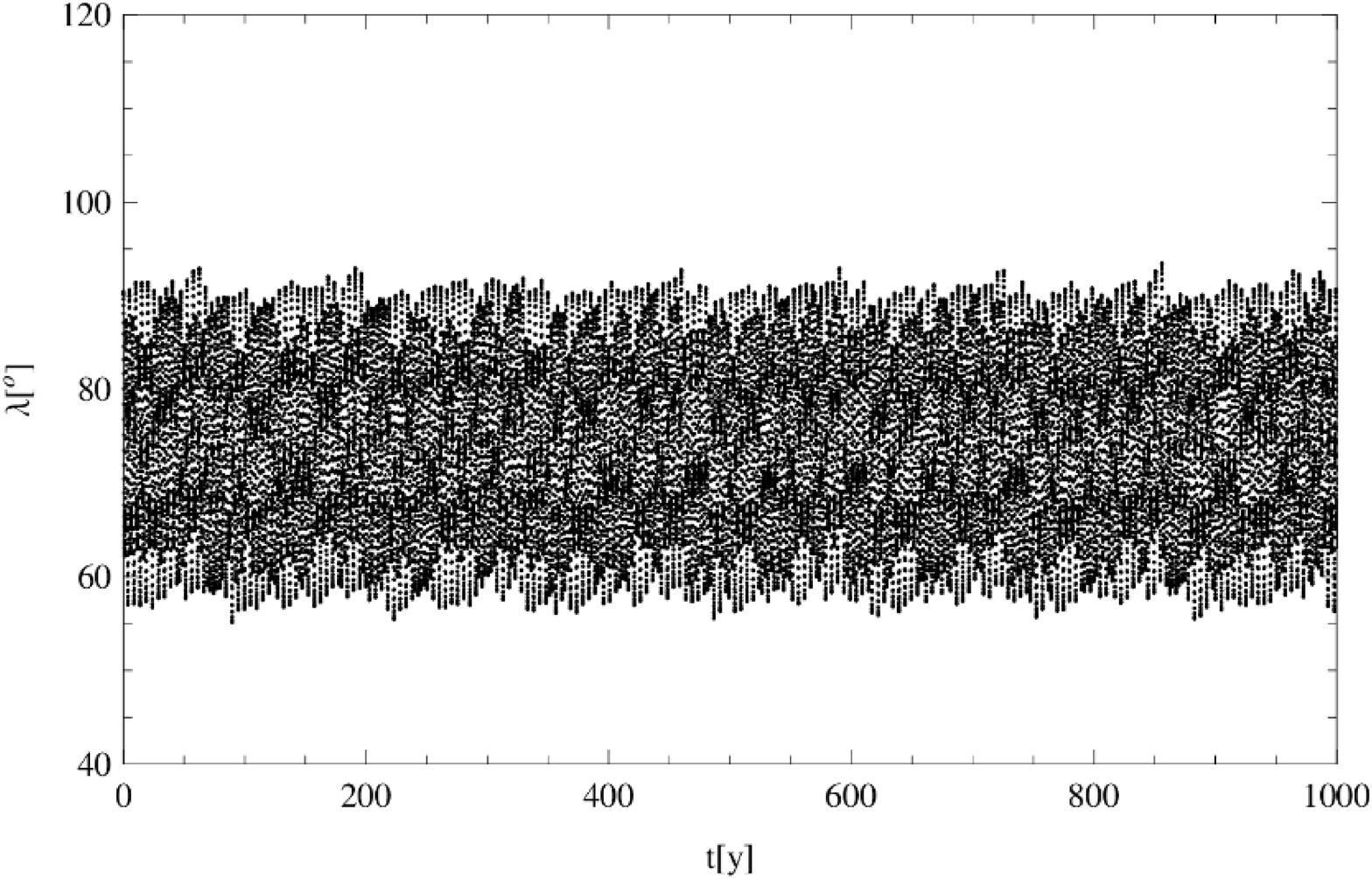}
\vglue0.5cm
\caption{Escape motion from the primary resonance (top) and trapped motion into primary resonance
(bottom). Variation of the semimajor axis $a$ and the resonant angle $\lambda$ for two orbits
in the libration region of primary resonance. Parameters and osculating initial conditions
are $A/m=15\ [mt^2/kg]$, $\eta=0$, $i(0)=10^o$, $e=0.2$, $\omega(0)=10^o$, $\Omega(0)=20^o$
and: $a(0)=42170\, km$, $\lambda(0)=45^o$ (or $M(0)=15^o$) (upper panels);  $a(0)=42190\, km$,
$\lambda(0)=90^o$  (or $M(0)=60^o$) (bottom panels). Compare with the right panel of Figure~\ref{fig:FLI_Am1_Am15}. }
\label{fig:Am15}
\end{figure}

\begin{figure}
\centering
\vglue0.1cm
\hglue0.1cm
\includegraphics[width=.45\linewidth]{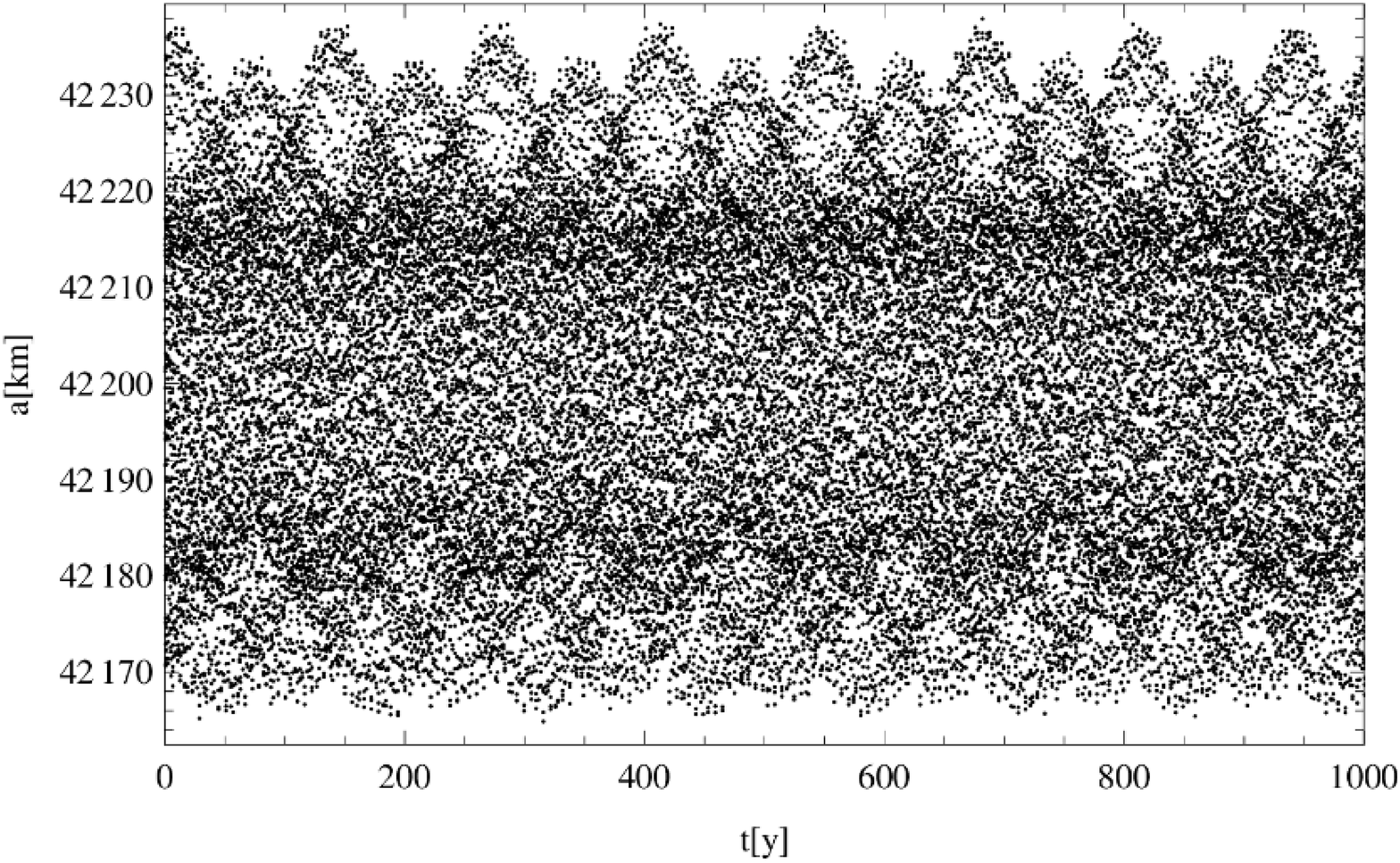}
\includegraphics[width=.45\linewidth]{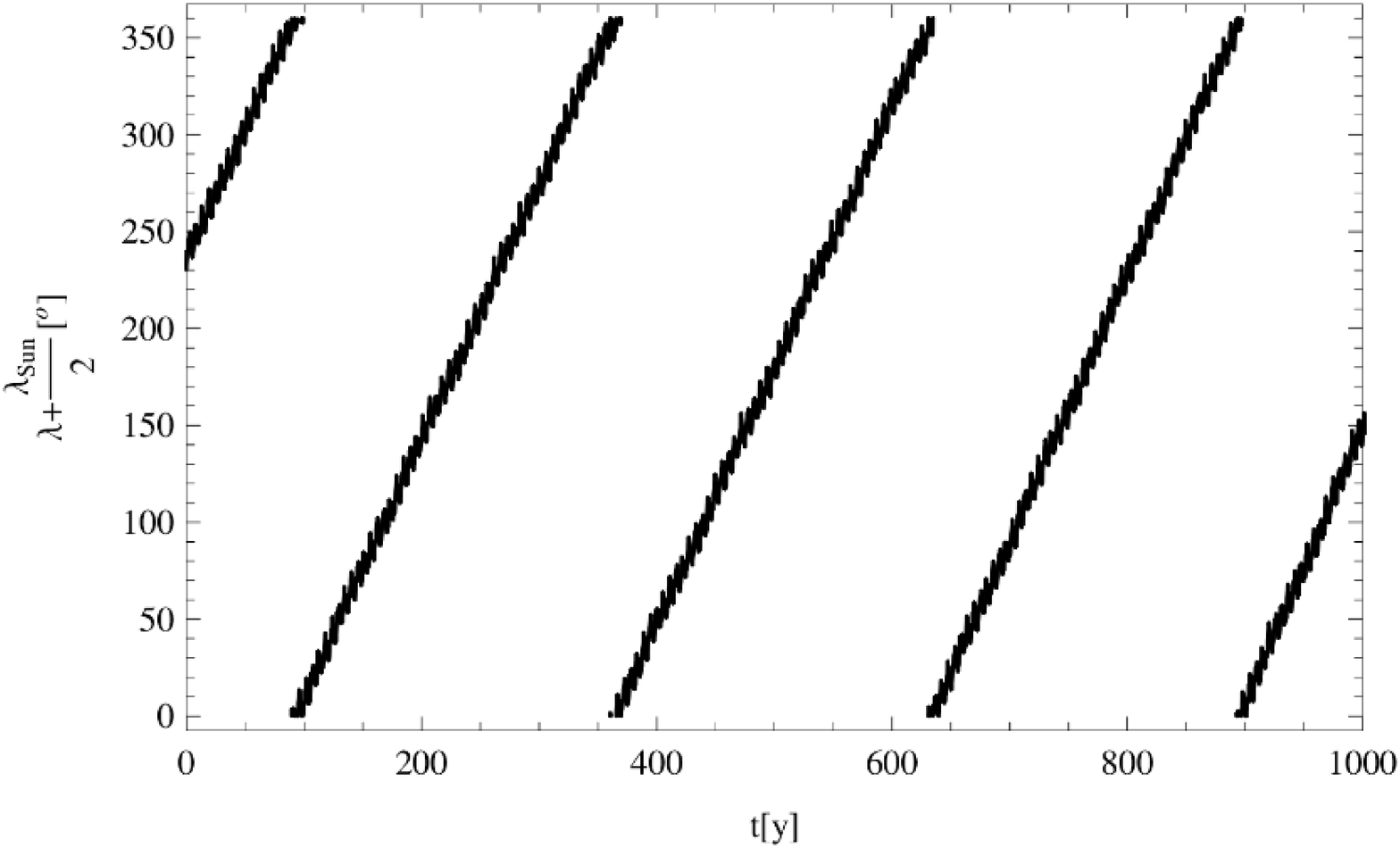}\\
\includegraphics[width=.45\linewidth]{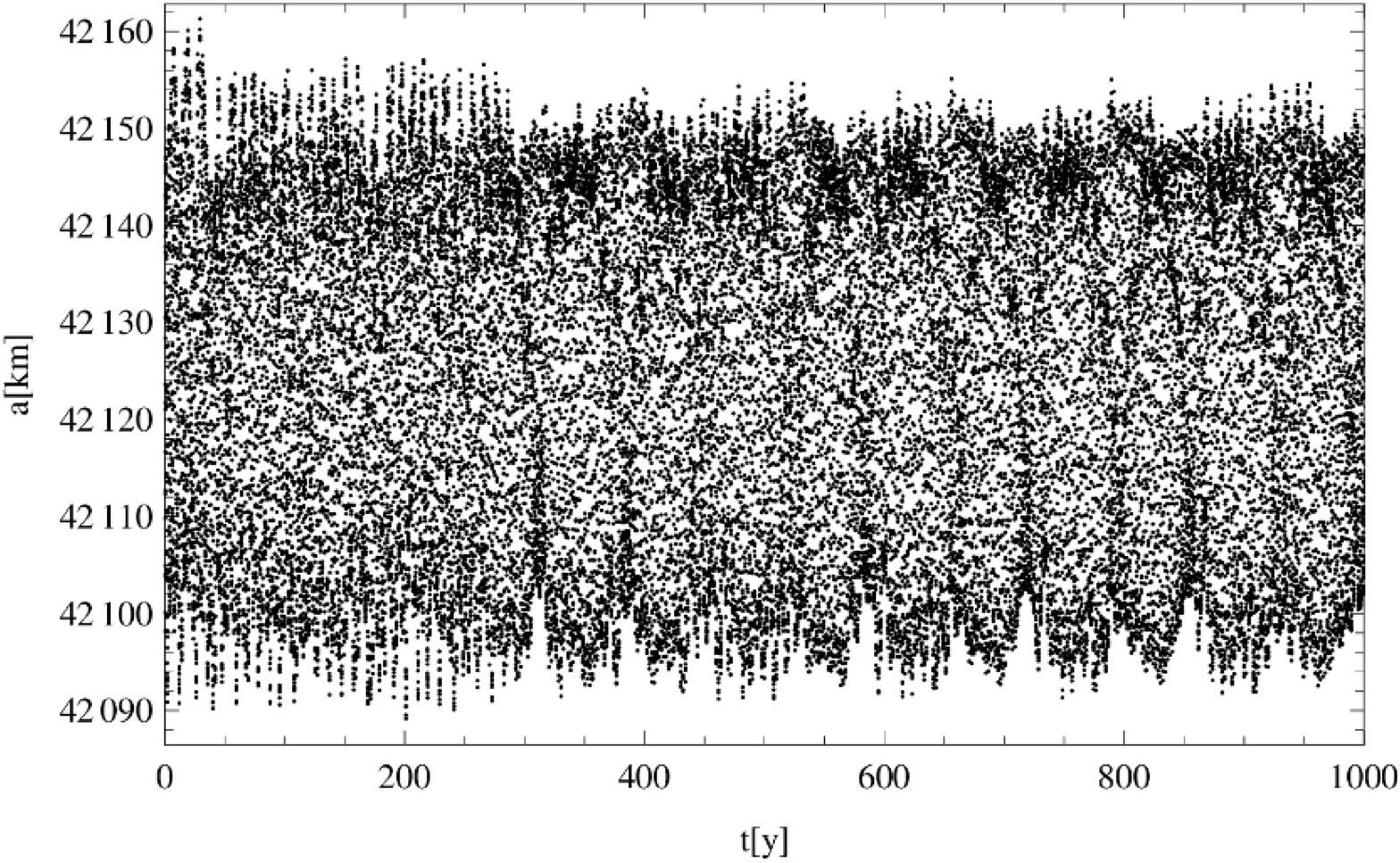}
\includegraphics[width=.45\linewidth]{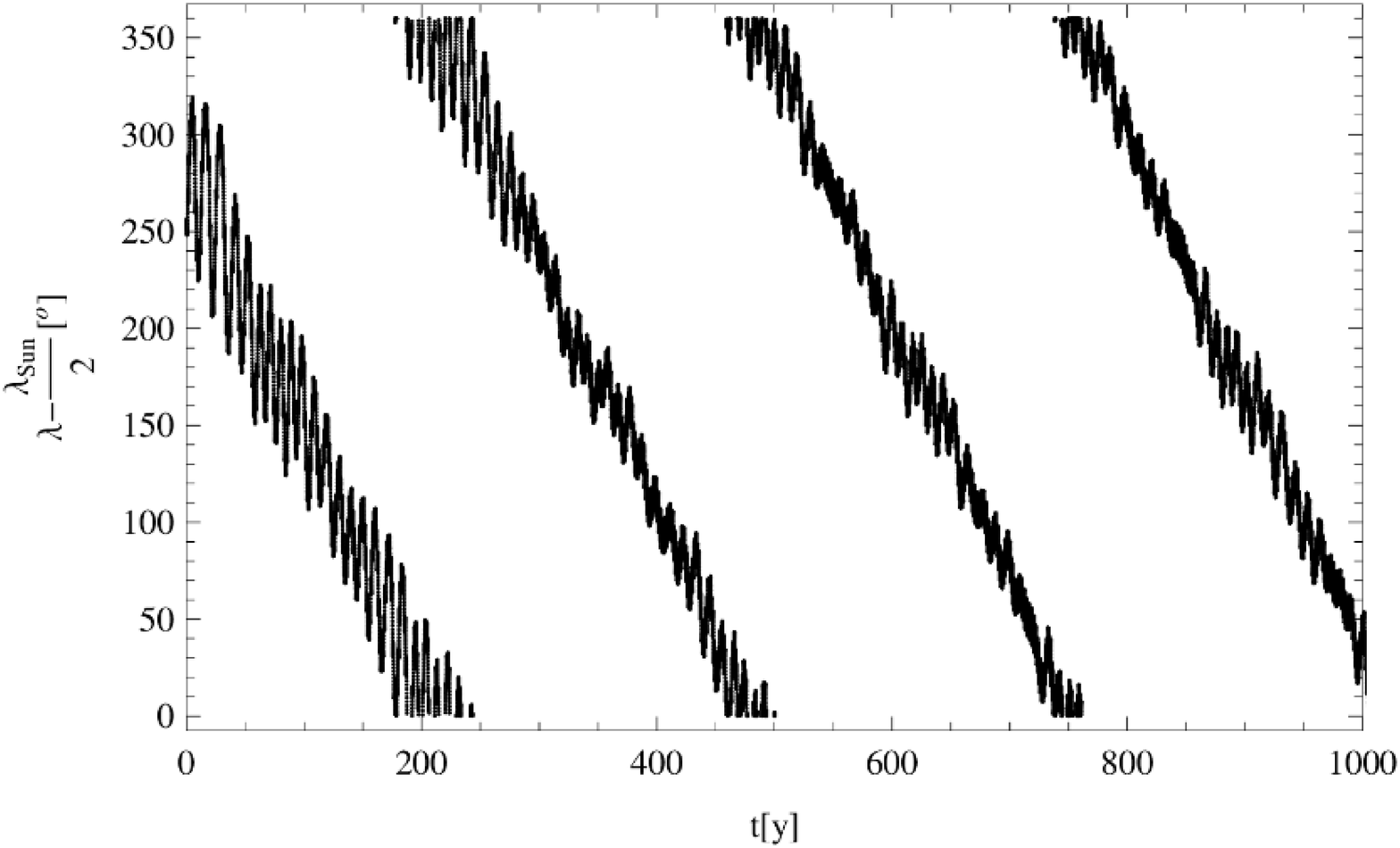}
\vglue0.5cm
\caption{Trapped motions into secondary resonances
$\dot{\lambda}+\frac{\dot{\lambda}_{Sun}}{2}=0$ (top) and
$\dot{\lambda}-\frac{\dot{\lambda}_{Sun}}{2}=0$ (bottom).  Variation of the
semimajor axis $a$ and the resonant angle $\lambda+\lambda_{Sun}/2$ (top) and
$\lambda-\lambda_{Sun}/2$ (bottom) for two orbits trapped in the secondary
resonances $\dot{\lambda}+\frac{\dot{\lambda}_{Sun}}{2}=0$ (top) and
$\dot{\lambda}-\frac{\dot{\lambda}_{Sun}}{2}=0$ (bottom).  Parameters and
osculating initial conditions are $A/m=15\ [mt^2/kg]$, $\eta=0$, $i(0)=10^o$,
$e=0.2$, $\omega(0)=10^o$, $\Omega(0)=20^o$ and: $a(0)=42190\, km$,
$\lambda(0)=-130^o$ (or $M(0)=-160^o$) (upper panels);  $a(0)=42100\, km$,
$\lambda(0)=-110^o$ (or $M(0)=-140^o$) (bottom panels).  Compare with the right
panel of Figure~\ref{fig:FLI_Am1_Am15}.}
\label{fig:Am15_secondary_resonances}
\end{figure}

\begin{figure}
\centering
\vglue0.1cm
\hglue0.1cm
\includegraphics[width=.45\linewidth]{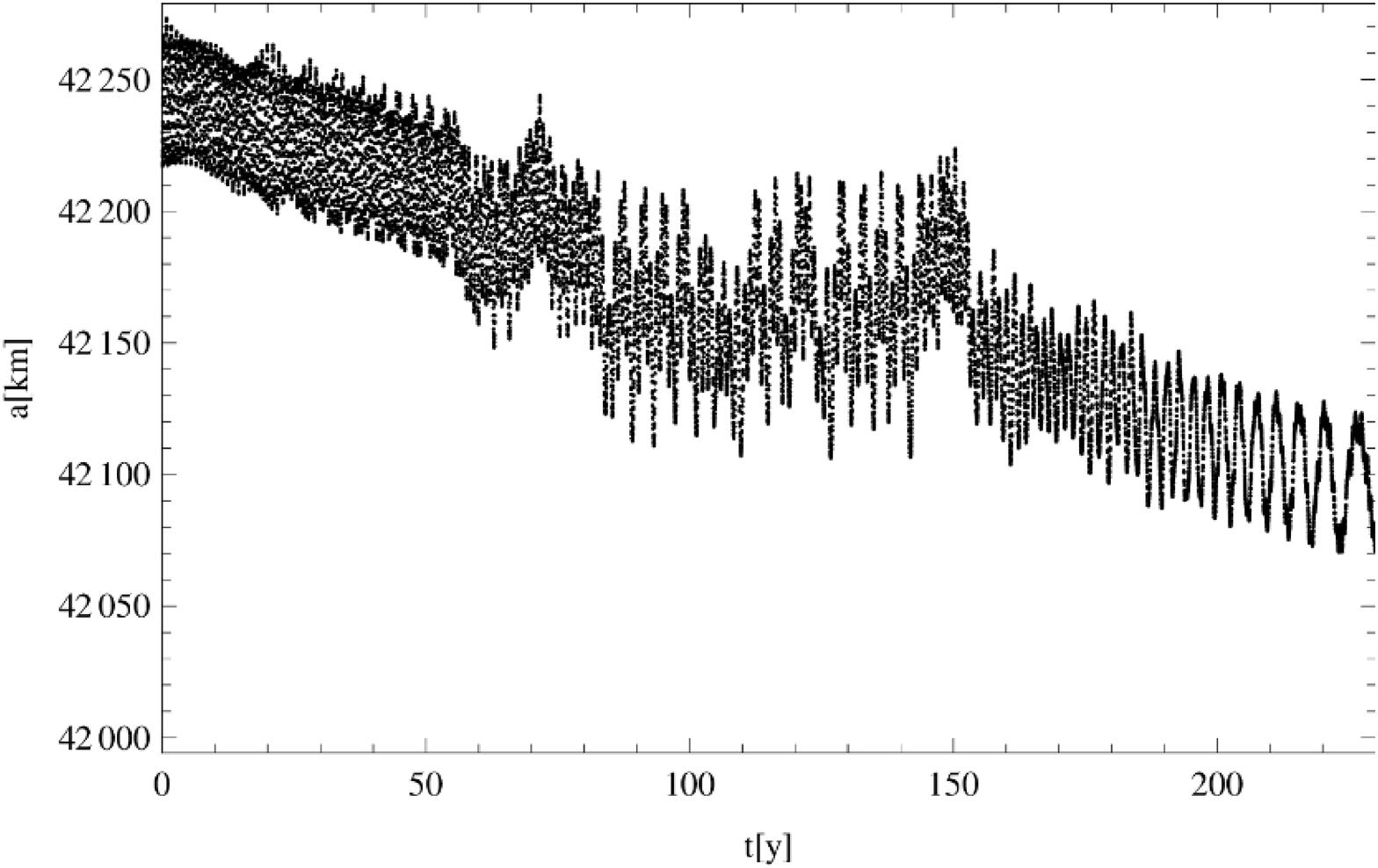}
\includegraphics[width=.45\linewidth]{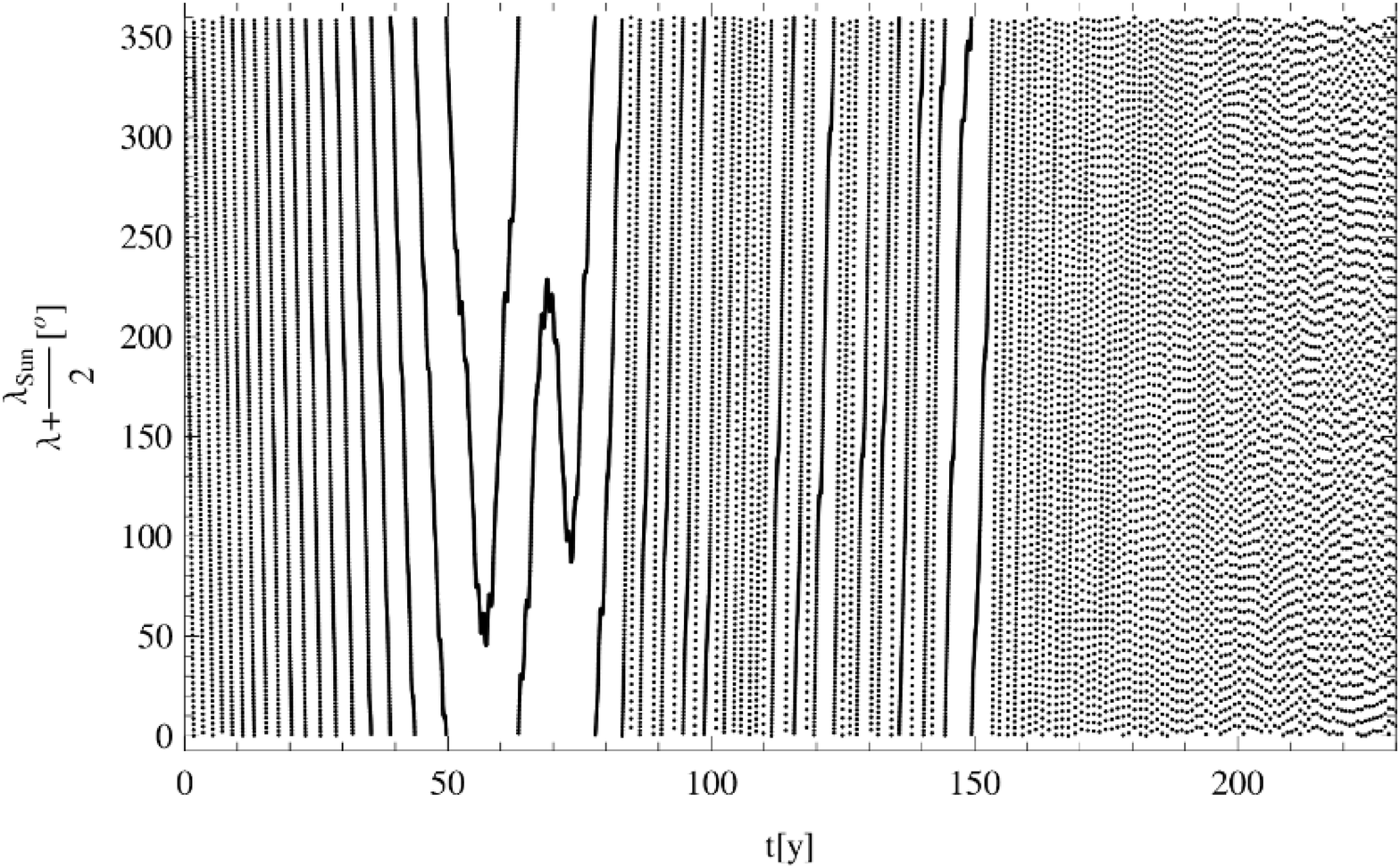}\\
\includegraphics[width=.45\linewidth]{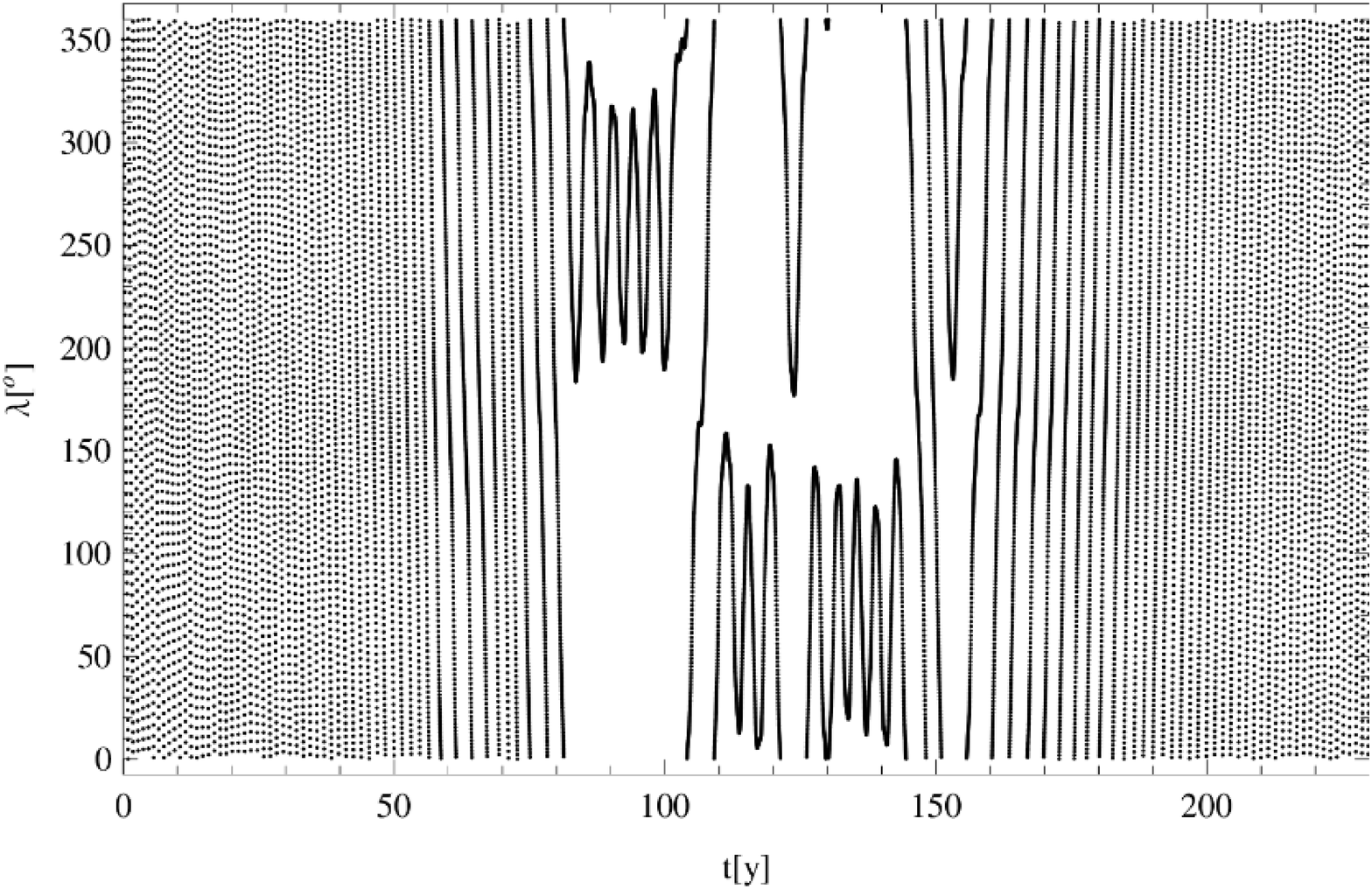}
\includegraphics[width=.45\linewidth]{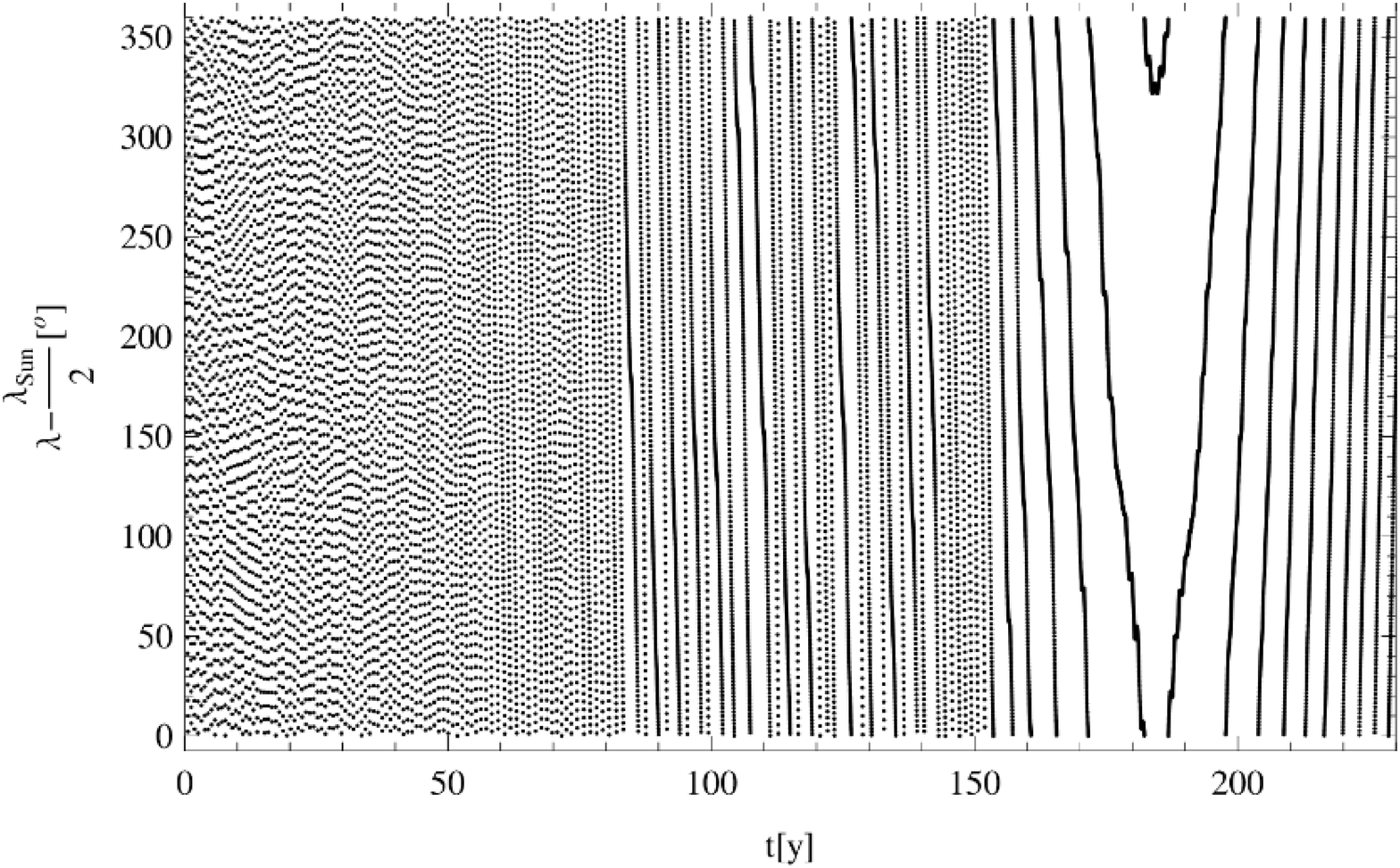}
\vglue0.5cm
\caption{Initial conditions are taken outside (above) the GEO 1:1 resonance.
Variation of the semimajor axis $a$ (top left) and the angles
$\lambda+\lambda_{Sun}/2$ (top right), $\lambda$ (bottom left),
$\lambda-\lambda_{Sun}/2$ (bottom right)  for an orbit having the following
parameters and osculating initial conditions: $A/m=15\ [mt^2/kg]$, $\eta=0$,
$i(0)=10^o$, $e=0.2$, $\omega(0)=10^o$, $\Omega(0)=20^o$,  $a(0)=42233\, km$
and $\lambda(0)=-5^o$ (or $M(0)=-35^o$). Compare with the right panel of
Figure~\ref{fig:FLI_Am1_Am15}.}
\label{fli:Am15_outside}
\end{figure}

However, when the area-to-mass ratio parameter is larger, say $A/m=15\,
[mt^2/kg]$, then besides the libration region associated to the primary
resonance $\dot{\lambda}=0$, some new structures are visible in the right panel
of Figure~\ref{fig:FLI_Am1_Am15}, which account for some secondary resonances
involving a linear combination of the geostationary resonant angle $\lambda$
with the longitude of the Sun $\lambda_{Sun}$ (see \cite{Valketal2009, Lem2009,
CellettiGales2015} for a detailed description of the web of secondary
resonances appearing outside the geostationary resonance as a product of the
interaction of solar radiation pressure with different tesseral resonances). In
fact, as Figures~\ref{fig:Am15}, \ref{fig:Am15_secondary_resonances} and
\ref{fli:Am15_outside} infer, the six larger libration islands visible  in the
right panel of Figure~\ref{fig:FLI_Am1_Am15} are due to the following
resonances: $\dot{\lambda}+{1\over 2} \dot{\lambda}_{Sun}=0$ (the two blue
regions located on top of the plot), $\dot{\lambda}=0$ (the two largest
islands) and $\dot{\lambda}-{1\over 2} \dot{\lambda}_{Sun}=0$ (the regions located at
the bottom of the plot). The chaotic region that surrounds the libration
islands is due to the interaction of these three resonances.

Comparing the patterns shown in the two panels of
Figure~\ref{fig:FLI_Am1_Am15}, we notice that a larger area-to-mass ratio, say
$A/m=15\, [mt^2/kg]$ (right plot), strongly deforms the phase space plots, as a
result of the influence of the short periodic part of the disturbing forces, in
particular as an effect of the action of the solar radiation pressure.  Each
disturbing force, namely the oblateness of the Earth, the attraction of the
Moon and of the Sun, and the solar radiation pressure, induces a short periodic
variation of the orbital elements.  These elements, used in the framework of
the Cartesian formulation and called \sl osculating \rm orbital elements,
differ from the \sl mean \rm orbital elements used in the secular theories. For
perturbations relatively small in magnitude, the difference between the mean
and osculating elements is not so evident. However, for large perturbations,
like that due to the effect of solar radiation pressure when $A/m=15\,
[mt^2/kg]$, there is a remarkable difference between the osculating and mean
elements.  However, we underline that in the rest of this section we deal with the
osculating orbital elements.

Within the conservative dynamical background described above, let us now
consider the dissipative effects induced by the PR/SW--drag.  By merging the
results given by the secular theory with numerical investigations, in the
following we describe  the weak influence of PR/SW drag and exemplify with some
concrete examples the complex dynamical behavior near the GEO 1:1 resonance on
large time scales.

Let us recall first that the stability analysis presented in
Section~\ref{subsec:equilibria} shows that the equilibria of the simplified
resonant dissipative system \eqref{ressys} are repellors. So, as theories of
dynamical systems suggest, the initial conditions located in the vicinity of these
points do not evolve toward but rather away from them. Thus, within the
framework of the dissipative system, the libration regions in
Figure~\ref{fig:FLI_Am1_Am15} should become a sort of ``basins of repulsion".
However, as Figure~\ref{sta} shows, the positive real parts of the eigenvalues
$(\zeta_1, \zeta_2)$ of the linearized system are very small in comparison
with the absolute values of the imaginary parts.  Therefore, in numerical
investigations we expect this effect to be very small even on long time scales.
Because the PR/SW-drag effect is weak, we will still use the terminology
``libration regions", even in the case of the full (dissipative) model, and not
``basins of repulsion" as we should normally adopt in the framework of
dissipative dynamical systems.

\begin{figure}
\centering
\vglue0.1cm
\hglue0.1cm
\includegraphics[width=.45\linewidth]{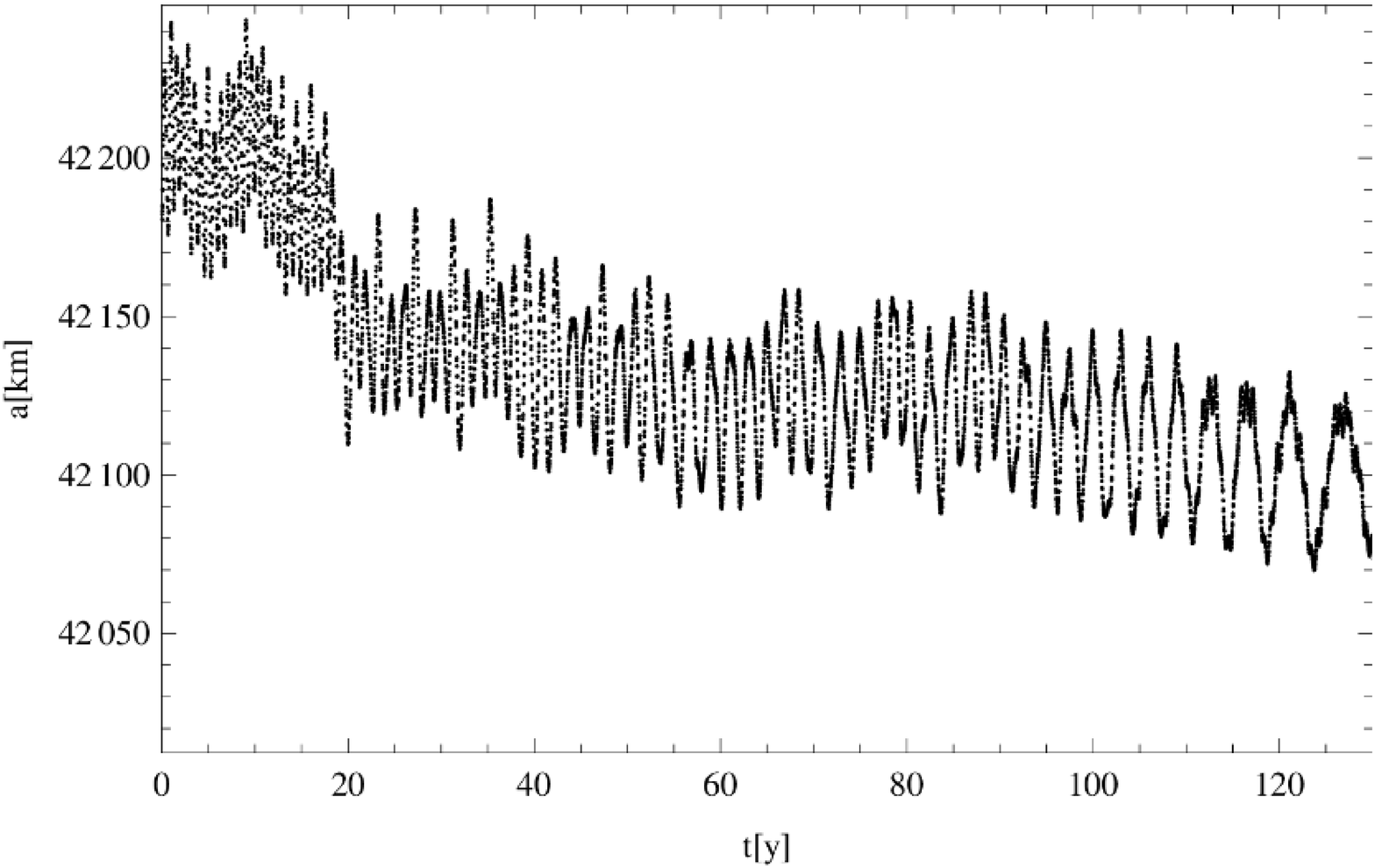}
\includegraphics[width=.45\linewidth]{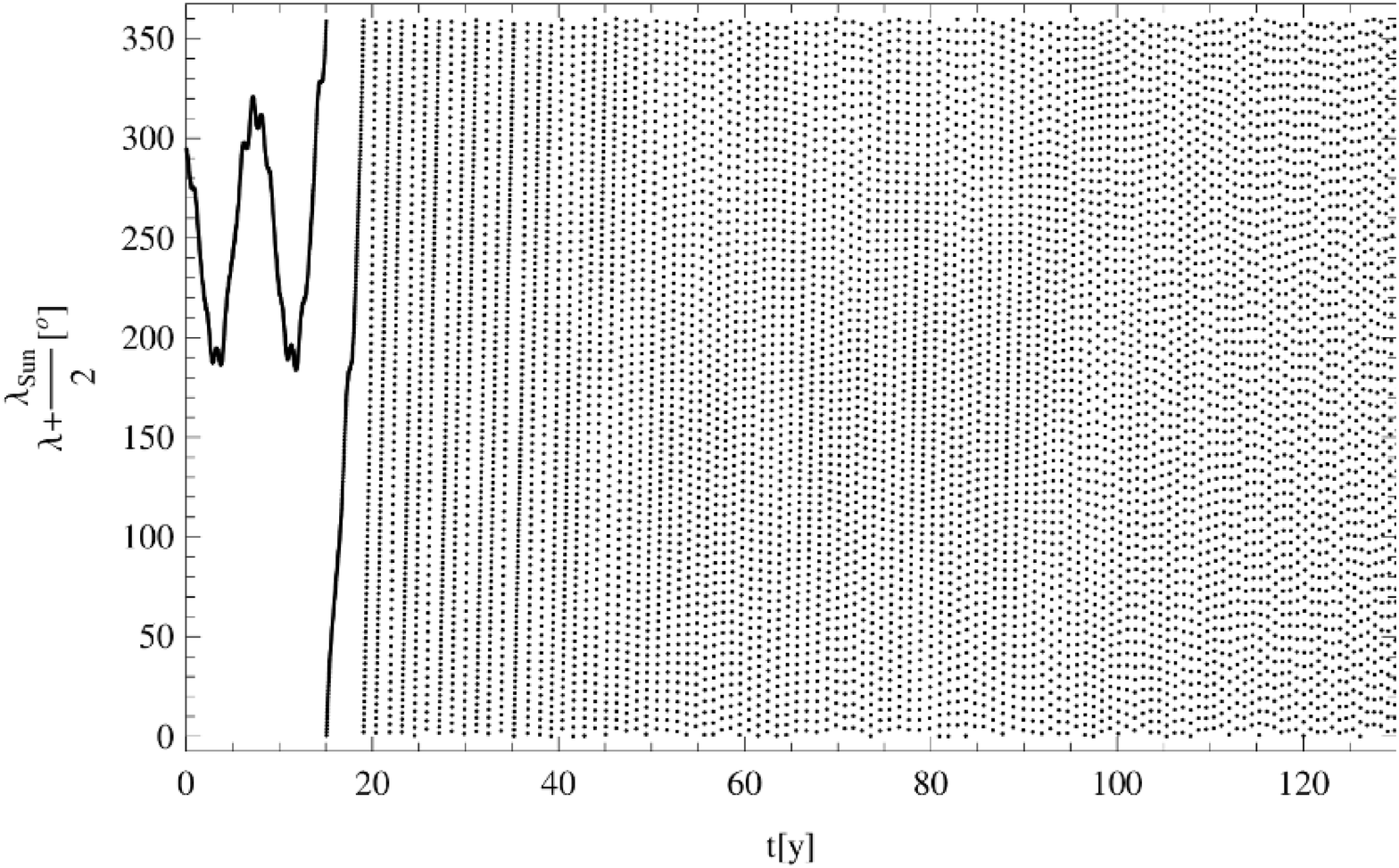}\\
\includegraphics[width=.45\linewidth]{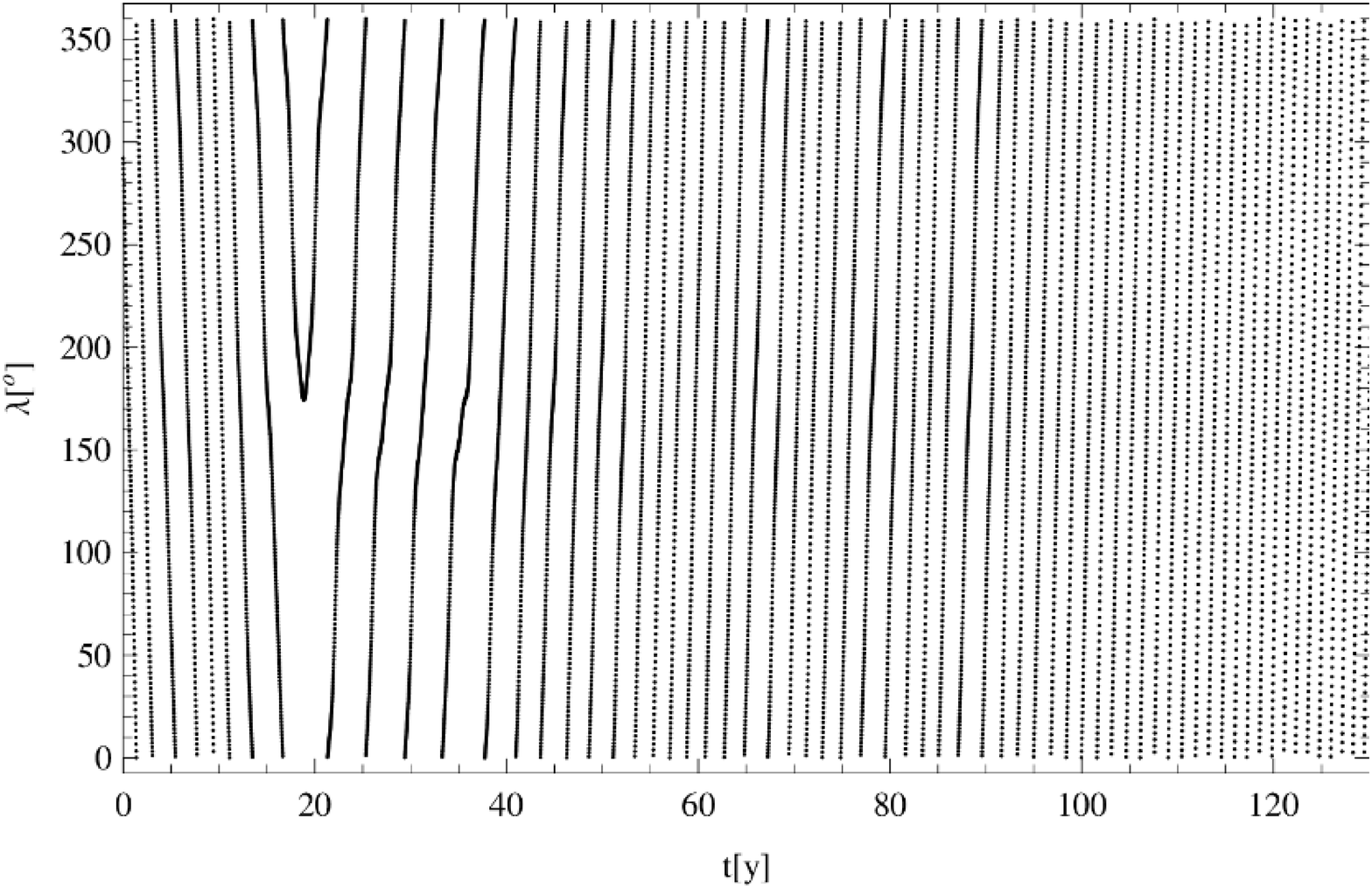}
\includegraphics[width=.45\linewidth]{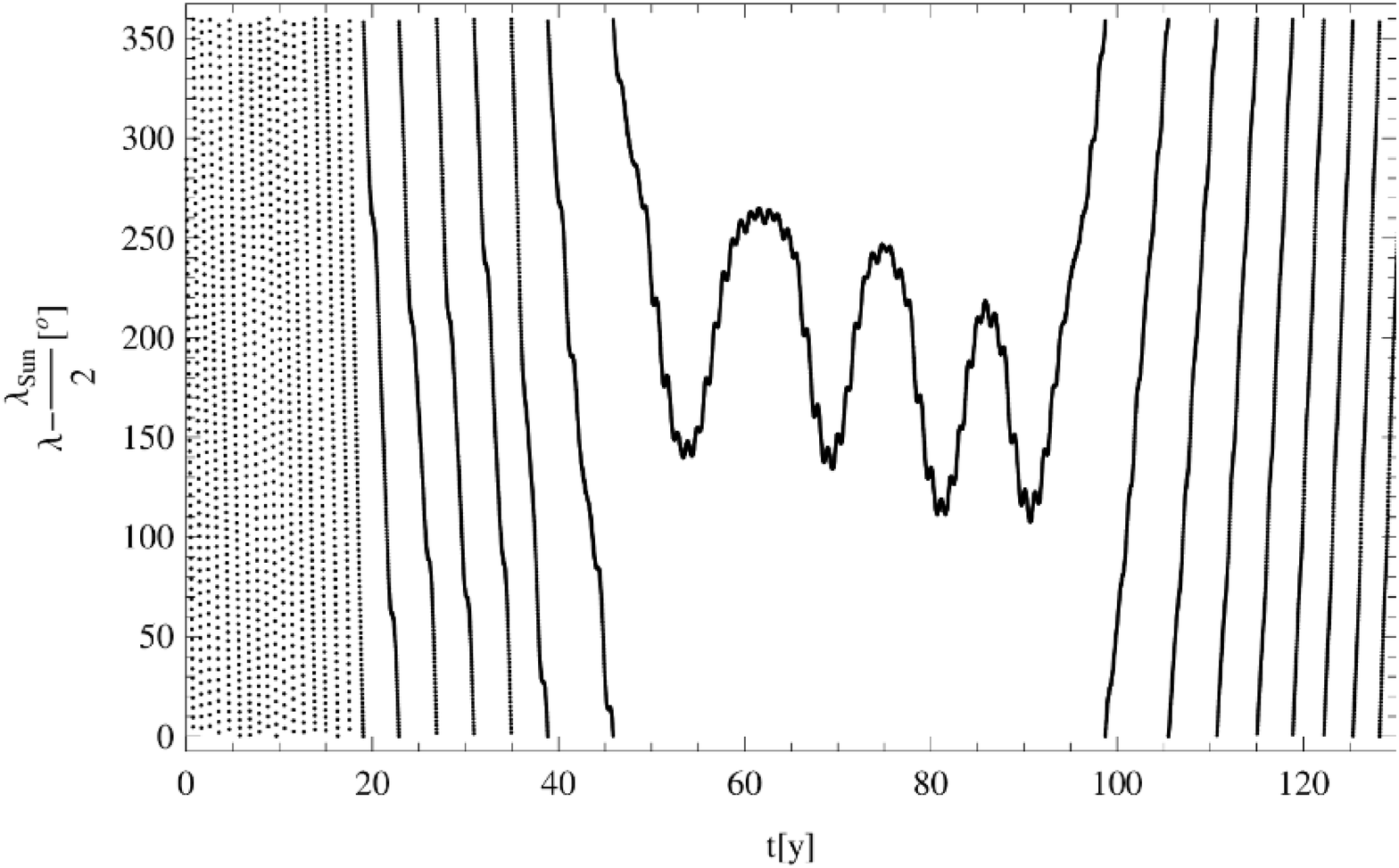}
\vglue0.5cm
\caption{Initial conditions are taken within the chaotic region surrounding the
GEO 1:1 resonance.  Variation of the semimajor axis $a$ (top left) and the
angles $\lambda+\lambda_{Sun}/2$ (top right), $\lambda$ (bottom left),
$\lambda-\lambda_{Sun}/2$ (bottom right)  for an orbit having the following
parameters and osculating initial conditions: $A/m=15\ [mt^2/kg]$, $\eta=0$,
$i(0)=10^o$, $e=0.2$, $\omega(0)=10^o$, $\Omega(0)=20^o$,  $a(0)=42178\, km$
and $\lambda(0)=-50^o$ (or $M(0)=-80^o$). Compare with the right panel of
Figure~\ref{fig:FLI_Am1_Am15}.}
\label{fli:Am15_chaotic_region}
\end{figure}

In Figures~\ref{fig:Am1}, \ref{fig:Am15}, \ref{fig:Am15_secondary_resonances},
\ref{fli:Am15_chaotic_region} and \ref{fli:Am15_outside} we report some results
obtained by propagating several osculating initial conditions for a time
reaching at most 1000 years, starting from January 1.5, 2000 (J2000). All these
orbits are represented by green--black circles in
Figure~\ref{fig:FLI_Am1_Am15}, which provided a picture within the conservative
case.

The combined influence of the dissipative effects (the fact that equilibrium
points are repellors) and of the conservative part can lead to {\it escape
motions} after a transient time, as shown by the top panels of
Figure~\ref{fig:Am1} for $A/m=1\, [mt^2/kg]$, and the top panels of
Figure~\ref{fig:Am15} for  $A/m=15\, [mt^2/kg]$. However, the initial condition
should be close enough to the separatrix for $A/m=1\, [mt^2/kg]$, or
sufficiently near the chaotic region for $A/m=15\, [mt^2/kg]$ (see
Figure~\ref{fig:FLI_Am1_Am15}).  Besides, the escape time is very large, more
than 900 years in the case of the orbit analysed in the top panels of
Figure~\ref{fig:Am1} for $A/m=1\, [mt^2/kg]$. These initial conditions lead to
an escape orbit; however, our tests
show that a small change in initial conditions, for example in $\lambda$, from
$\lambda(0)=5^o$ to $\lambda(0)=6^o$, leads to a {\it trapped motion} for more
than 1000 years. In addition, we find drift,  temporary capture and release
from resonance for $\lambda=20{}^o$ (middle panels of Figure~\ref{fig:Am1}),
but also drift and long-term capture (bottom panels of Figure~\ref{fig:Am1})
for $\lambda=0{}^o$.

For larger area-to-mass ratios, the effects induced by both
the solar radiation pressure and PR/SW--drag increase in intensity.
Figure~\ref{fig:Am15}, top panels, obtained for $A/m=15\, [mt^2/kg]$, show an
escape orbit characterized by a smaller escape time; due to large
perturbations, it crosses multiple dynamical regimes (temporary escapes and
captures), before it escapes definitively at about 310 years. In fact, as the
numerical results are showing, the escape time depends on multiple factors:
parameters, how close the initial conditions are from the separatrix (or from
the chaotic region), the magnitude of the perturbing forces, the interaction
between various perturbations.


For $A/m=15\, [mt^2/kg]$, given the fact that the dynamical background is much
more complex than for relatively small area-to-mass ratios, the matter is a
little bit more complicated.  The orbit described by the top plots of
Figure~\ref{fig:Am15}, and which is located in the
libration region of the primary resonance (compare with the right panel of
Figure~\ref{fig:FLI_Am1_Am15}), is an escape orbit. This does not mean that any
orbit located in libration regions is an escape orbit. As discussed above, it
depends on how close the initial condition is from the equilibrium  point
located inside the libration region. For example, the bottom plots of
Figure~\ref{fig:Am15} describe a temporarily trapped motion into a primary
resonance. It is interesting to point out that trapped motions exist also into
the secondary resonances $\dot{\lambda}+{1\over 2} \dot{\lambda}_{Sun}=0$ and
$\dot{\lambda}-{1\over 2}\dot{\lambda}_{Sun}=0$, as shown in
Figure~\ref{fig:Am15_secondary_resonances}.

In order to have a holistic picture of the dynamics for $A/m=15 \, [mt^2/kg]$,
inside the GEO 1:1 resonance,  we should describe the behavior of the orbits located in the
chaotic region. Since the interaction between the primary resonance and the
secondary resonances is large, an initial condition from the chaotic region leads to an
escape orbit on a scale of time of the order of tens (at most two hundred) years.
To reveal the complex interplay between resonances and PR/SW--drag effect,
in Figure~\ref{fli:Am15_chaotic_region} and \ref{fli:Am15_outside} we represent the evolution of the
semimajor axis $a$, and the resonant angles $\lambda+\lambda_{Sun}/2$, $\lambda$
and $\lambda-\lambda_{Sun}/2$, associated to each resonance. As
Figures~\ref{fli:Am15_chaotic_region} and \ref{fli:Am15_outside} show, the orbit is affected by
all resonances, since there is a temporary trapping in each resonance
at different intervals of time. The escape time is  less than 100 years.

Finally, to completely describe the behavior in the near vicinity of the GEO 1:1 resonance,
one should understand the behavior of the orbits located initially at a longer distance
than that at which the resonance pattern is located. As effect of the interaction
between the dissipative and conservative parts of the perturbations, numerical simulations
show that there are possible {\it jumps}, usually for small area-to-mass ratios
(not shown here), {\it temporary captures into the primary resonance}
(middle panels of Figure~\ref{fig:Am1}), {\it temporary captures into primary and
secondary resonance} (Figures~\ref{fli:Am15_chaotic_region} and \ref{fli:Am15_outside}) or even {\it captures} for
times longer than 1000 years (bottom panels of Figure~\ref{fig:Am1}). It is important
to note that in either of the cases, capture into primary resonance or capture into
secondary resonances, the orbit does not approach toward the center of libration island,
but rather remains on a long time scale in a very close neighborhood of the separatrix.







\begin{figure}
\centering
\vglue0.1cm
\hglue0.1cm
\includegraphics[width=.3\linewidth]{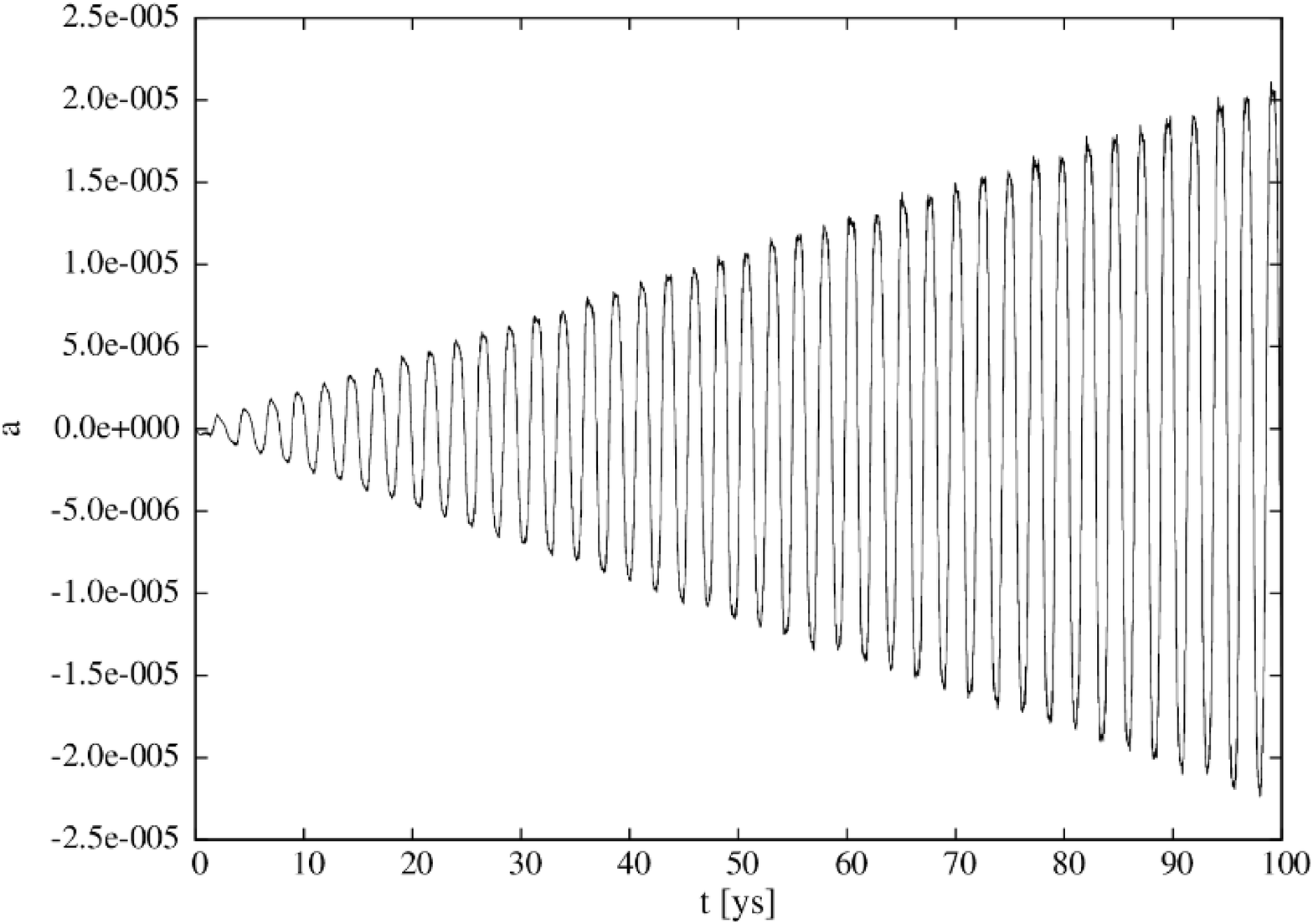}
\includegraphics[width=.3\linewidth]{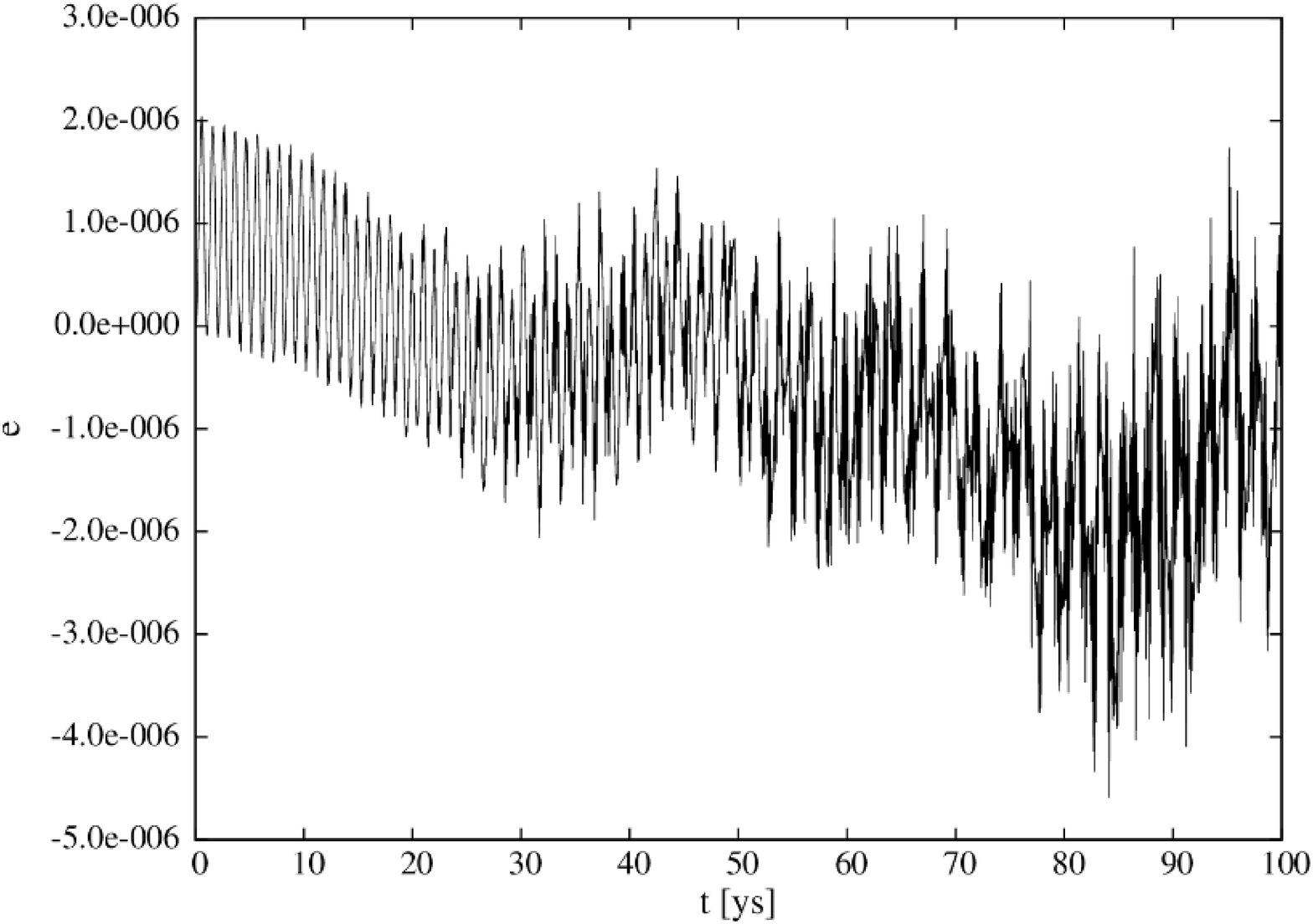}
\includegraphics[width=.3\linewidth]{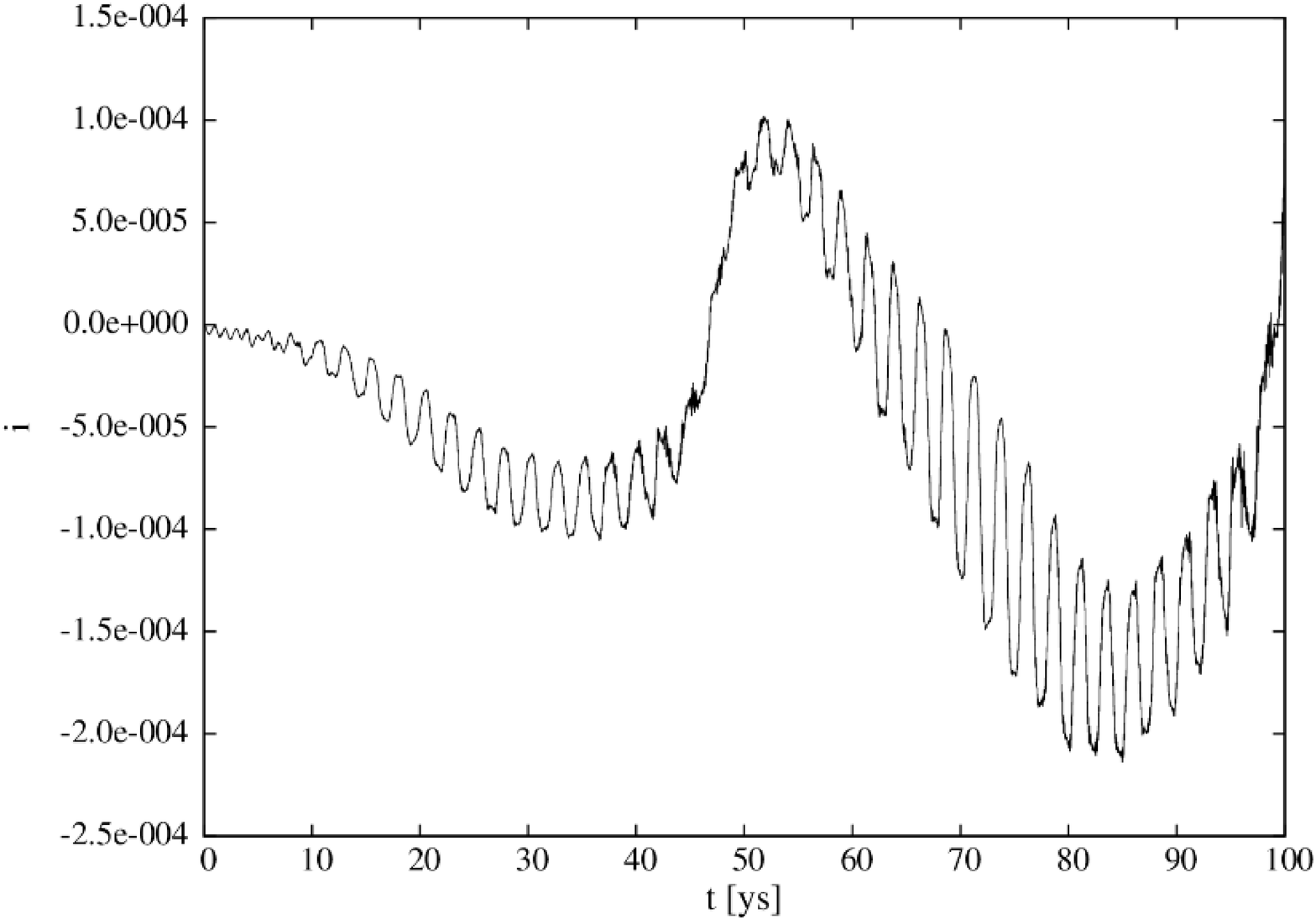}\\
\includegraphics[width=.3\linewidth]{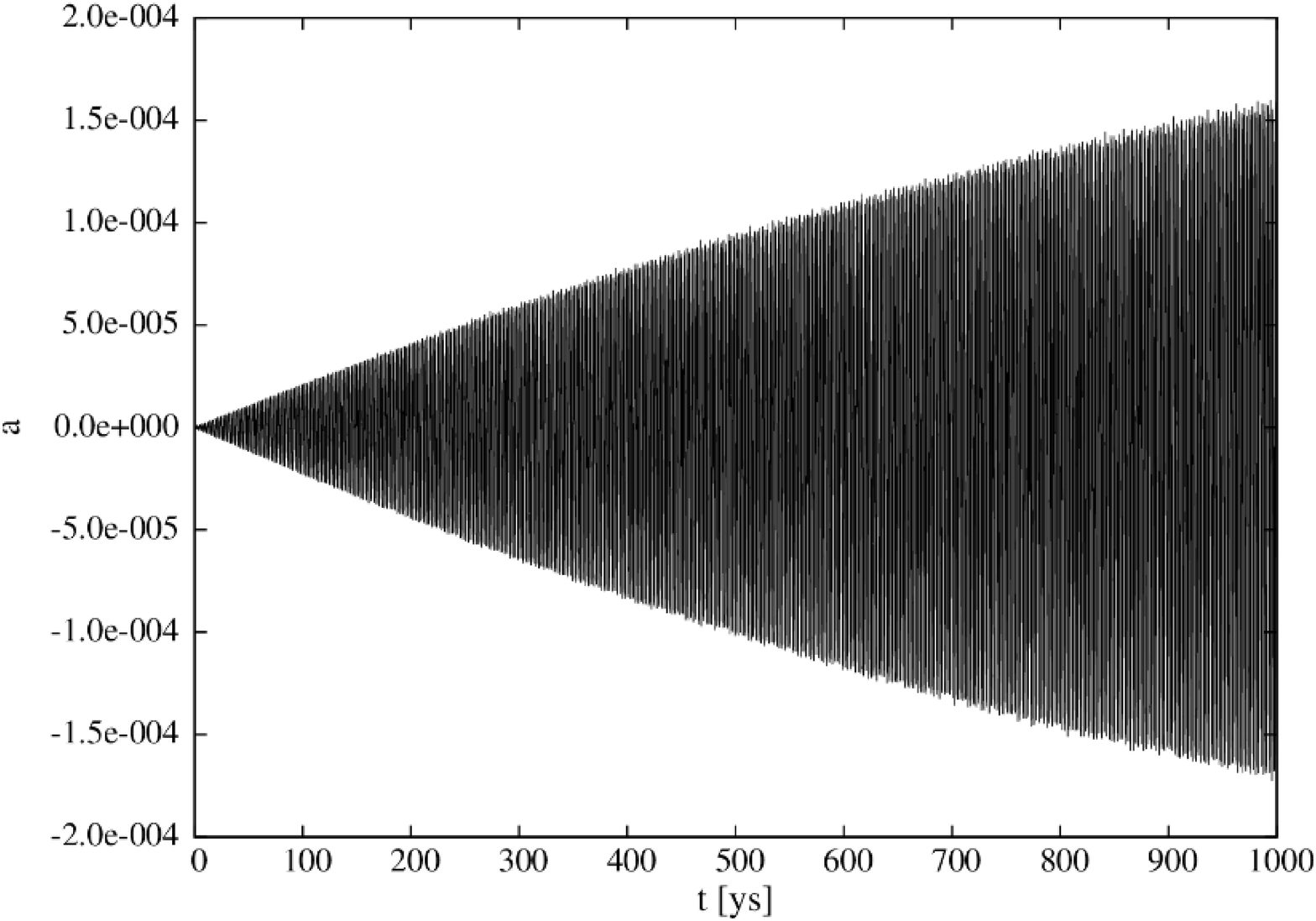}
\includegraphics[width=.3\linewidth]{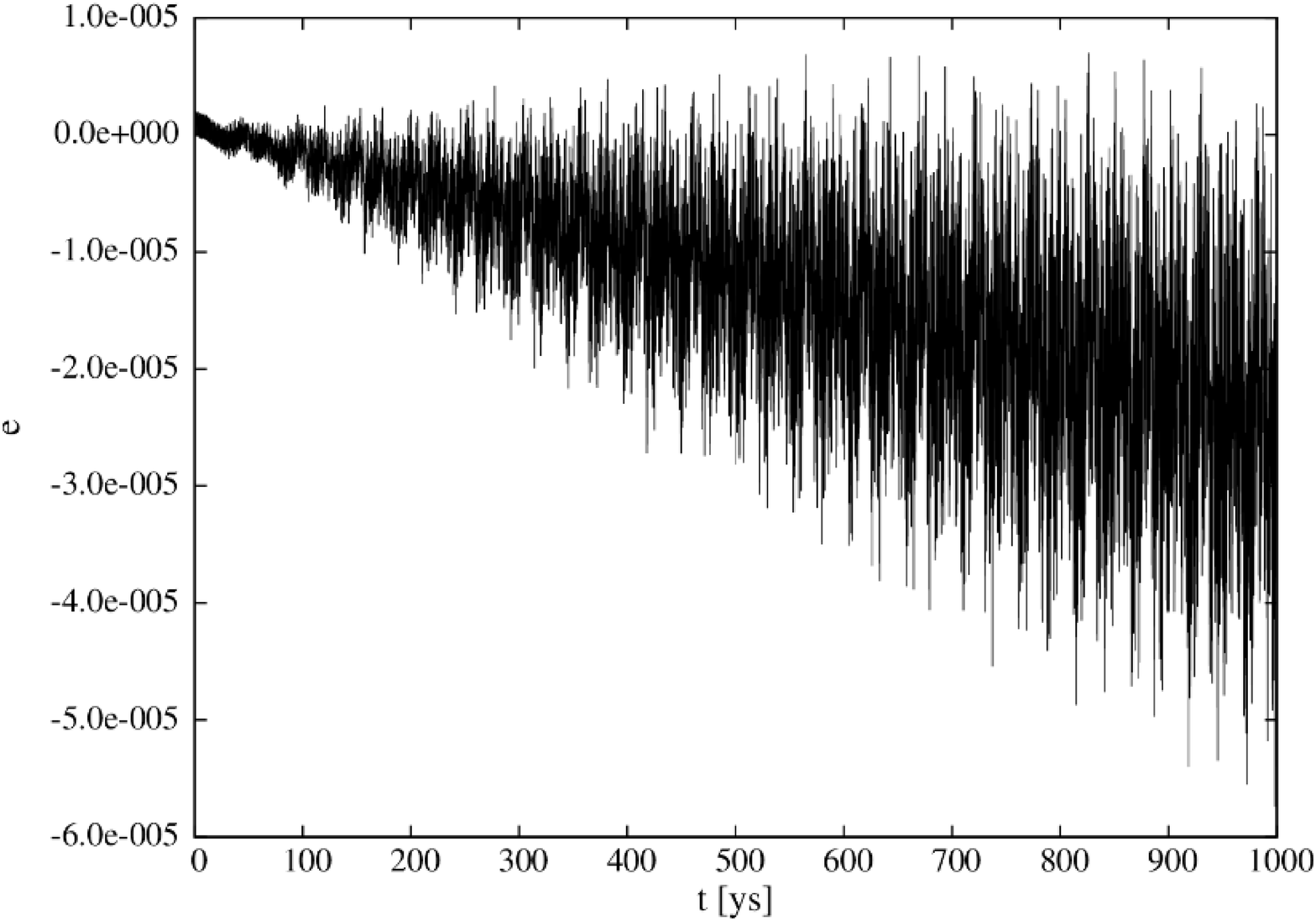}
\includegraphics[width=.3\linewidth]{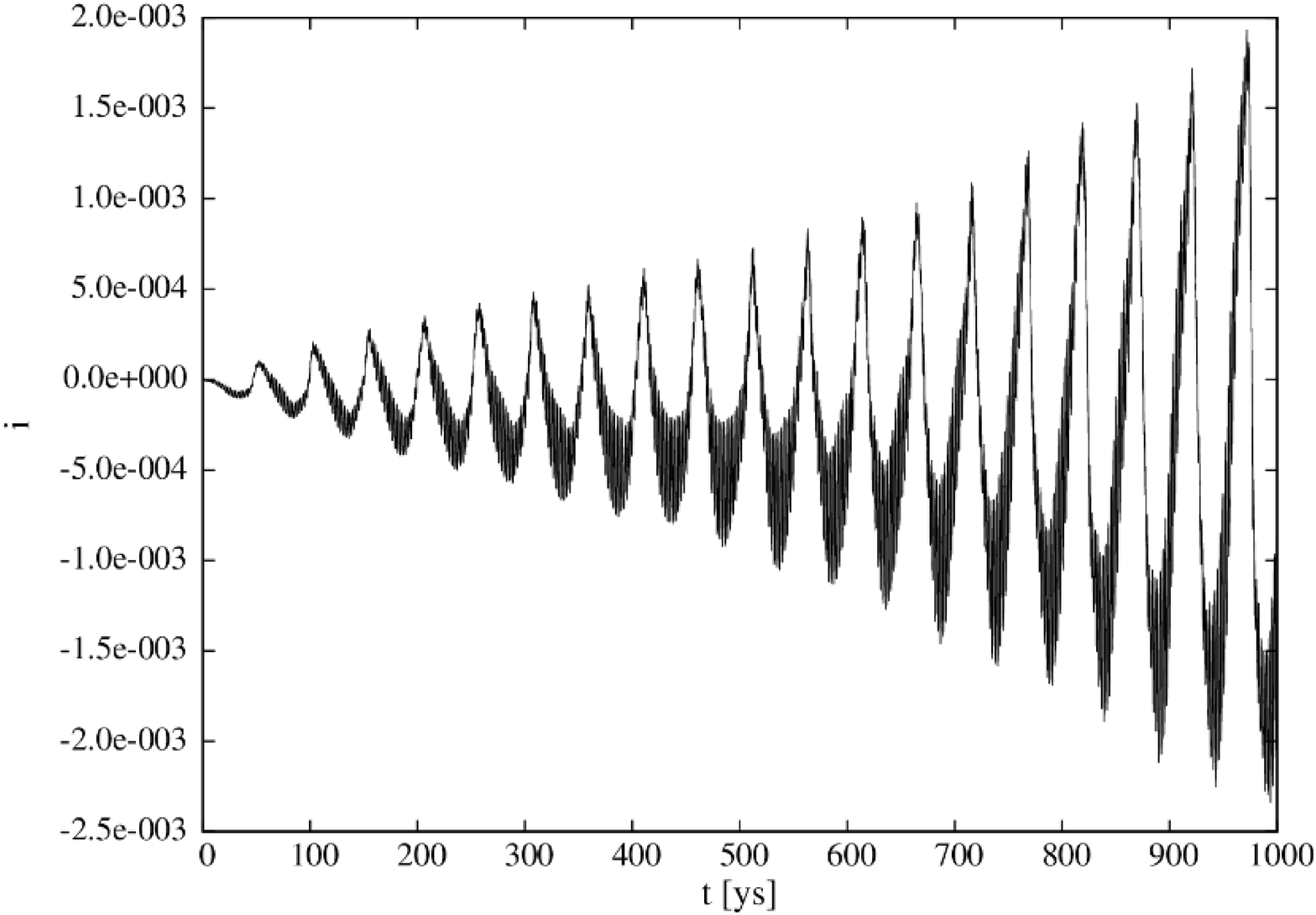}
\vglue0.5cm
\caption{Difference of the integration with and without Poynting--Robertson
effect. The other effects included in the equations of motion are: Earth's
harmonics up to $n=m=3$, SRP, Sun and Moon.  Here we take $a=42164.8$ km,
$e=0.1$, $i=3^o$, $\omega=5^o$, $\Omega=10^o$, $M=20^o$, $A/m=1\ [mt^2/kg]$.  The left
panels refer to the semimajor axis (in units of $a_{geo}\equiv 42\,164.17\
[km]$, the middle panels to the eccentricity, the right panels to the
inclination (in degrees).  The upper panels show the integration over 100
years, while the lower panels over 1000 years.}
\label{GEOtime}
\end{figure}

As a final experiment, we perform a numerical integration of the equations of motion
with and without PR effect. Figure~\ref{GEOtime} shows the difference in orbital elements
$(a,e,i)$ over 100 and 1000 years. The longer is the timescale, the stronger is the
influence of the PR effect. We also computed several other integrations, all showing
that PR effect is  not so relevant for small values of the initial eccentricity, as well as
for small values of the area-to-mass ratio.

\section{Conclusions}\label{sec:conclusions}
It is well known that Poynting-Robertson and solar wind drag contribute as dissipative effects
acting over long time scales on artificial satellites and space debris.
We explore the behavior of the dynamics in the vicinity of the GEO 1:1 resonance for two sample
objects characterized by $A/m=1 \, [mt^2/kg]$ and $A/m=15 \, [mt^2/kg]$. Given the fact that the population
of objects in the GEO region continues to grow, it is important to know how small objects
(equivalently, large objects having high area-to-mass ratios), which could result from a
catastrophic event, behave under the combined effect of dissipative forces and
conservative perturbations. Understanding their dynamics is crucial for risk evaluation
and mitigation strategies.

Although several studies were performed in the past years to evaluate the role of the dissipative effects due to PR/SW,
we are not aware of any investigation of PR/SW drag combined with a careful analysis of the dynamical properties
of the phase space. We know that equilibria
are repellors both within and outside the librational region, but the evolution of a space debris strongly depends on
its initial location as well as its area-to-mass ratio. Indeed, the study performed in the previous sections show that a debris with
relatively small area-to-mass ratio, say $A/m=1\ [mt^2/kg]$, might undergo different behaviors,
even when changing a little the initial conditions. The three case studies considered in
Figure~\ref{fig:FLI_Am1_Am15}, left panel, show that, although being all close to the separatrix, the debris can be
trapped into resonance for a long time and then escape (Figure~\ref{fig:Am1}, top panels), or
rather it can be only temporary trapped (Figure~\ref{fig:Am1}, middle panels), or trapped
after a transient time (Figure~\ref{fig:Am1}, bottom panels).

On the other hand, when the area-to-mass ratio is large, say $A/m=15\ [mt^2/kg]$, the structure
of the phase space changes significantly: chaotic motions
occupy a relevant portion of the phase space, while secondary resonances make their appearance, thus reflecting
the interaction with the longitude of the Sun (see Figure~\ref{fig:FLI_Am1_Am15}, right panel).

The investigation given in this work leads to a zoo of dynamical behaviors under PR/SW drag: temporary trapping or escape from the primary resonance,
temporary trapping or escape from one of the secondary resonances.
All these information can be conveniently used to monitor the long-term behavior of space debris, and in
particular can be used to evaluate the decay rate due to PR/SW drag. Since the dynamical behavior is very sensitive to
small displacements of the object in the phase space, one could even design a strategy which exploits 
PR/SW drag to move space debris within different regions for the safeguard of operational satellites.


\begin{appendices}

\section*{Appendix A}

\subsection*{Gravity field}

The quantities $P_{nm}$ in \equ{eq1} are defined in terms of the standard
Legendre polynomials:
\beqno
P_n(x)\equiv {1\over {2^n n!}}\ {{d^n}\over {dx^n}}\{(x^2-1)^n\}\ ,\qquad
P_{nm}(x)\equiv (1-x^2)^{m\over 2}\ {{d^m}\over {dx^m}}\{P_n(x)\} \ .
\eeqno
The coefficients $C_{nm}$, $S_{nm}$ are defined as
\beqano
C_{nm}&\equiv&{{2-\delta_{0m}}\over m_E}\
{{(n-m)!}\over {(n+m)!}}\ \int_{V_E} ({r_p\over R_E})^n\
P_{nm}(\sin\phi_p)\
\cos(m\lambda_p)\delta(\underline{r}_p)\ dV_E\nonumber\\
S_{nm}&\equiv&{{2-\delta_{0m}}\over m_E}\ {{(n-m)!}\over
{(n+m)!}}\ \int_{V_E} ({r_p\over R_E})^n\ P_{nm}(\sin\phi_p)\
\sin(m\lambda_p)\delta(\underline{r}_p)\ dV_E\ ,
\eeqano
where $(r_p,\lambda_p,\phi_p)$ are the spherical coordinates of some
point $P$ inside the Earth ($\delta_{jm}$ is the Kronecker symbol). \\

\subsection*{Rotation matrices}

The rotation matrices are defined as follows:

\beqano
{\mathfrak R}_1(\varphi)=\left(
\begin{array}{ccc}
 1 & 0 & 0 \\
 0 & \cos\varphi & -\sin\varphi \\
 0 & \sin\varphi & \cos\varphi \\
\end{array}
\right) \ , \quad
{\mathfrak R}_3(\varphi)=\left(
\begin{array}{ccc}
 \cos\varphi & -\sin\varphi & 0 \\
 \sin\varphi & \cos\varphi & 0 \\
 0 & 0 & 1 \\
\end{array}
\right) \ .
\eeqano

\subsection*{Osculating orbital elements}

The orbital elements $a$, $e$, $i$, $\omega$, $\Omega$, $M$ are
obtained from the Cartesian position $\vec r$ and velocity $\vec v$ 
by means of the following procedure. Let $\mu=\G m_E$ and 
$\vec{I}=(1,0,0)$, $\vec{J}=(0,1,0)$, $\vec{K}=(0,0,1)$. Furthermore,
$r=|\vec{r}|$, $v=|\vec{v}|$ and let $h \vec{K}=\vec{r}\wedge \vec{v}$ 
be the angular momentum with $h=|\vec{h}|$. From the eccentricity vector
$$
\vec{e}={{\vec{v}\wedge \vec{h}}\over {\mu}}-{{\vec{r}}\over {r}}
$$
we calculate the eccentricity $e=|\underline{e}|$. The semi-major axis is 
obtained by the relation
$$
h^2=\mu a(1-e^2)\ ,
$$
while the inclination can be obtained from
$$
\cos i={{\vec{h}\cdot \vec{K}}\over {h}}\ .
$$
Let $\vec{n} /  n$ be the unit nodal vector with $\vec{n}=\vec{K}\wedge \vec{h}$ and
$n=|\vec{n}|$; then, $\Omega$ is given by
$$
\cos\Omega= {\vec{n}\cdot \vec{I}\over n}\ ,\qquad
\sin\Omega= {\vec{n}\cdot \vec{J}\over n} \ .
$$
The argument of perigee is obtained from
$$
\cos\omega={\vec{n}\cdot \vec e\over {ne}}\ , \qquad
\sin\omega=({\vec{n}\over n}\wedge {\vec{e}\over e}) \cdot {\vec{h}\over h}\ .
$$
The true anomaly $f$ is given by
$$
\cos f={{\vec{e}\cdot \vec{r}}\over {e\, r}}\ ,
$$
while the eccentric anomaly is obtained from
$$
\cos E={{e+\cos f}\over {1+e\cos f}}\ ,
$$
which, together with Kepler's equation
$$
M=E-e\sin E\ ,
$$
provides the mean anomaly $M$.

\end{appendices}

\section*{Acknowledgements}

A.C. was partially supported by PRIN-MIUR 2010JJ4KPA$\_$009, GNFM/INdAM,
Stardust Marie Curie Initial Training Network, FP7-PEOPLE-2012-ITN, Grant
Agreement 317185. C.G. was supported by a grant of the Romanian National
Authority for Scientific Research and Innovation, CNCS - UEFISCDI, project
number PN-II-RU-TE-2014-4-0320.


\bibliographystyle{mn2e}
\bibliography{arxiv}{}

\label{lastpage}

\end{document}